\documentclass{aa}

\usepackage{mathbbol}
\usepackage[tbtags,fleqn]{amsmath}
\usepackage{amsfonts}
\usepackage{graphicx}
\usepackage{pstricks,pst-plot}
\usepackage{natbib}
\bibpunct{(}{)}{;}{a}{}{,}
\newcommand{\diff}{\mathrm d}

\newcommand{\vecbar}[1]{\bar{\vec #1}}

\DeclareMathOperator{\Cov}{Cov}


\begin{document}

\title{The noise of cluster mass reconstructions from a source
  redshift distribution}
\author{Marco Lombardi, Peter Schneider, and Carlos Morales-Merino}
\authorrunning{M. Lombardi et al.}
\titlerunning{The noise of cluster mass reconstructions from a source
  redshift distribution}
\institute{%
  Instit\"ut f\"ur Astrophysik und Extraterrestrische Forschung,
  Auf dem H\"ugel 71, D-53121 Bonn, Germany} 
\offprints{M. Lombardi}
\mail{lombardi@astro.uni-bonn.de}
\date{Received ***date***; accepted ***date***}
\abstract{%
  The parameter-free reconstruction of the surface-mass density of
  clusters of galaxies is one of the principal applications of weak
  gravitational lensing.  From the observable ellipticities of images
  of background galaxies, the tidal gravitational field (shear) of the
  mass distribution is estimated, and the corresponding surface mass
  density is constructed.  The noise of the resulting mass map is
  investigated here, generalizing previous work which included mainly
  the noise due to the intrinsic galaxy ellipticities.  Whereas this
  dominates the noise budget if the lens is very weak, other sources
  of noise become important, or even dominant, for the medium-strong
  lensing regime close to the center of clusters.  In particular, shot
  noise due to a Poisson distribution of galaxy images, and increased
  shot noise owing to the correlation of galaxies in angular position
  and redshift, can yield significantly larger levels of noise than
  that from the intrinsic ellipticities only.  We estimate the
  contributions from these various effects for two widely used
  smoothing operations, showing that one of them effectively removes
  the Poisson and the correlation noises related to angular positions
  of galaxies.  Noise sources due to the spread in redshift of
  galaxies are still present in the optimized estimator and are shown
  to be relevant in many cases.  We show how (even approximate)
  redshift information can be profitably used to reduce the noise in
  the mass map.  The dependence of the various noise terms on the
  relevant parameters (lens redshift, strength, smoothing length,
  redshift distribution of background galaxies) are explicitly
  calculated and simple estimates are provided.  \keywords{Dark matter
    -- gravitational lensing -- large-scale structure of Universe --
    Galaxies: clusters: general -- Methods: statistical} }

\maketitle


\section{Introduction}
\label{sec:introduction}

Weak gravitational lensing has been recognized as a powerful tool to
investigate the mass and mass distribution of clusters of galaxies
\citep{1985MNRAS.213..871W, 1990ApJ...349L...1T, 1990MNRAS.247..135K}.
In their pioneering paper \citet{KS} pointed out that the tidal
gravitational field of the cluster, as measured from the distortion of
image shapes of the faint background galaxy population, can be used to
reconstruct the two-dimensional projected mass profiles of clusters,
without referring to a parameterized mass model. This method, modified
in various ways later (see, e.g., \citealp{1995ApJ...439L...1K,
  1995A&A...302..639S, SS1, 1996ApJ...464L.115B, 1996ApJ...473...65S,
  SS3, 1998A&A...337..325S, 1998A&A...330..791L, 1998A&A...335....1L,
  RevBS} for a recent review) has been applied to more than a dozen
clusters up to now \citep[e.g.][]{1994ApJ...437...56F,
  1995ApJ...440..501S, 1996ApJ...469...73S, 1996ApJ...461..572S,
  1996A&A...314..707S, 1997ApJ...475...20L, 1998ApJ...497L..61C,
  1998ApJ...504..636H, 1999ARA&A..37..127M}.  One of the main results
from these studies is that in many clusters the projected mass
distribution closely follows the distribution of luminous cluster
galaxies.

Recently, a number of papers have been devoted to the study of the
noise properties of weak lensing mass reconstructions.  Such studies
are important to optimize the lensing techniques used.  For example,
\citet{1998A&A...335....1L} have provided estimates for the noise
properties of weak lensing mass maps, showing also that a particular
technique, previously described by \citet{1996A&A...305..383S}, leads
to optimal maps in the linear regime; analogously,
\citet{2000MNRAS.313..524W} has extended these results to maximum
likelihood estimators.

All these analytical studies, however, have been carried out using a
``mean field'' approximation, thus ignoring some contributions to the
noise.  In particular, Poisson noise and correlations on the positions
of source galaxies are not taken into account.  In this paper we
consider in detail the effects of these sources of noise and we
provide analytical estimates for them.  We also show that, in many
situations, these extra contributions to the noise are non-negligible.
In particular, as already pointed out by \citet{2001ApJ...546...47B},
the noise due to correlations of the source galaxies \textit{does
  not\/} decrease with the density of observed galaxies, and thus puts
an \textit{intrinsic\/} limit on the accuracy of weak lensing mass
maps.  On the other hand, we also show that the use of some redshift
information can significantly reduce the noise due to the correlation
of galaxies and, at least partially, the Poisson noise.

The paper is organized as follows.  We briefly describe standard weak
lensing mass reconstruction techniques in
Sect.~\ref{sec:local-shear-estim}.  In
Sect.~\ref{sec:statistical-analysis} we introduce the statistical
methods used to study the properties of the shear and the mass maps
obtained in weak lensing reconstructions, and we apply those methods
to obtain some general results.  A very useful approximation technique
is discussed in Sect.~\ref{sec:slowly-varying-fields}.  This technique
allows us to obtain simple analytical results and to study the impact
of various sources of noise on the shear estimation.  A more accurate
shear estimator, which leads to significantly smaller noise, is
studied in Sect.~\ref{sec:more-accurate-shear}.
Section~\ref{sec:impact-observations} is devoted to the impact of
observations on reconstructions of weak lenses.  If information on the
source redshifts is available, then some sources of noise can be
significantly reduced.  Two different estimators that can take
advantage of redshift information are discussed in
Sect.~\ref{sec:case-source-with}.  Finally, the main results obtained
in this paper are summarized in Sect.~\ref{sec:conclusions}.  In order
to simplify the discussion and to avoid long calculations, we have
moved several derivations to the appendices.  In particular, in
Appendix~\ref{sec:linear-approximation-1} we recall a standard
statistical technique; in Appendices~\ref{sec:deta-stat-analys} and
\ref{sec:slowly-vary-fields-1} we carry out the statistical analysis
for the estimator of Sect.~\ref{sec:more-accurate-shear}; in
Appendix~\ref{sec:calc-estim-with} we sketch the calculations for the
estimators of Sect.~\ref{sec:case-source-with}.

The weak lensing approximation is taken to be valid in this paper, so
that the shear is taken as an observable [see below Eq.~\eqref{eq:1}].
The Einstein-de~Sitter cosmological model is assumed and the Hubble
constant is parameterized as $H_0 = 100h \mbox{ km s}^{-1} \mbox{
  Mpc}^{-1}$.

\section{Local shear estimators}
\label{sec:local-shear-estim}

In the weak lensing limit, the observed ellipticity $\epsilon^{(n)}$
of a galaxy ($n \in \{1, 2, \dots, N\}$ denotes the galaxy considered)
is related to the true source ellipticity $\epsilon^{\mathrm{s}(n)}$:
\begin{equation}
  \label{eq:1}
  \epsilon^{(n)} = \epsilon^{\mathrm{s}(n)} + Z^{(n)} \gamma \bigl(
  \vec\theta^{(n)} \bigr) \; ,
\end{equation}
where $\gamma(\vec\theta)$ is the lens shear field for fiducial
sources at infinite redshift, $Z^{(n)} = Z\bigl( z^{(n)} \bigr)$ is
the \textit{cosmological weight\/} for the $n$\nobreakdash-th galaxy,
and $\vec\theta^{(n)}$ is its observed position on the sky.  We recall
that $Z(z)$ depends on the assumed cosmological model, on the redshift
$z_\mathrm{d}$ of the lens, and on the redshift $z^{(n)}$ of the
galaxy.  More precisely, $Z(z)$ can be written as a ratio of angular
diameter distances between the observer and the lens, the observer and
the source, and the lens and the source \citep[see, e.g.,][]{RevBS}.

Equation~\eqref{eq:1} provides a simple way to infer the shear of a
lens.  Since source galaxies can be taken to be randomly oriented, so
that $\bigl\langle \epsilon^\mathrm{s} \bigr\rangle = 0$, the average
of galaxies close on angular position is a good estimator for the
shear.  The average, needed in order to obtain a high signal-to-noise
estimator from extremely noisy data (the source ellipticity), is
usually performed using a function $W(\vec\theta - \vec\theta')$ which
describes the \textit{relative\/} contribution of a galaxy located at
$\vec\theta'$ to the shear at the position $\vec\theta$.  Since only
relative values of $W$ are important, we can assume that this function
is normalized,
\begin{equation}
  \label{eq:2}
  \int_\Omega W(\vec\theta - \vec\theta') \, \diff^2 \theta' = 1 \; ,
\end{equation}
where $\Omega$ is the field in the sky where the observations are
made.  In principle, the normalization condition written above cannot
be satisfied for all points $\vec\theta$ of $\Omega$ unless $W$ is
constant.  In practice, if the area of observation $\Omega$ is large
compared to the scale of $W$, we can assume that Eq.~\eqref{eq:2}
holds for nearly all points $\vec\theta \in \Omega$, being false only
for the points close to the boundary of the field.  This
approximation, called \textit{framing} (see Lombardi \& Bertin 1998b),
greatly simplifies the analysis.  In the following, we will always use
this approximation.  Often, in weak lensing studies, a Gaussian weight
function is used:
\begin{equation}
  \label{eq:3}
  W(\vec\theta - \vec\theta') = \frac{1}{2 \pi \sigma_W^2} \exp \left(
    - \frac{\lvert \vec\theta - \vec\theta' \rvert^2}{2 \sigma_W^2}
  \right) \; .
\end{equation}

Using the \textit{smoothing function\/} $W$, two simple \textit{unbiased\/}
estimators for the shear can be provided.  The first one is the
traditional \citet{KS} estimates for the shear:
\begin{equation}
  \label{eq:4}
  \hat\gamma(\vec\theta) = \dfrac{1}{\langle Z \rangle \rho} 
  \sum_{n=1}^N \epsilon^{(n)} W \bigl( \vec\theta - \vec\theta^{(n)}
  \bigr) \; .
\end{equation}
Here, $\rho$ is the density of observed galaxies, for simplicity taken
to be constant on the whole field of observation (as a reference
value, a density of $30$ galaxies per square arcmin can be easily
reached today).  In this equation and in the whole paper we will use
the hat ($\hat{\ }$) to denote \textit{estimated\/} or
\textit{measured\/} quantities.  Equation~\eqref{eq:4} is very simple,
but has the disadvantage of being significantly affected by Poisson
noise.  In fact, it is reasonable to expect some variations of the
\textit{local\/} density of galaxies in the field.  In such cases, we
would have additional noise on $\gamma$ simply due to changes on
$\rho$.

As argued by \citet{SS2}, since the positions of galaxies are readily
available, we can modify the estimator into
\begin{equation}
  \label{eq:5}
  \hat\gamma(\vec\theta) = \dfrac{\sum_{n=1}^N \epsilon^{(n)}
    W \bigl( \vec\theta - \vec\theta^{(n)} \bigr)}{\langle Z \rangle
    \sum_{n=1}^N W \bigl( \vec\theta - \vec\theta^{(n)} \bigr)} \; .
\end{equation}
We will refer to this estimator as \textit{balanced}, because it, in
contrast to \eqref{eq:4}, explicitely takes into account the local
number density of source galaxies.  Equation~\eqref{eq:5} is actually
often used in weak lensing studies \citep[see,
e.g.,][]{2000A&A...363..401L}.

In this paper we will study in detail both estimators and obtain their
expected variances.  We will consider initially the estimator
\eqref{eq:4}, which is much easier to study than \eqref{eq:5}.
Although non-optimal, this estimator is very important for many
reasons: (i) It is the first one considered in the literature, in the
classical paper by \citet{KS}; (ii) it is simpler to study and hence
allows exact analytical results; (iii) it has been used in the
literature \citep[e.g.][]{1994ApJ...437...56F, 1995ApJ...446L..55T}
and actually is still used by some authors \citep{1997AJ....113..521F,
  1997AJ....114...14F, 1997ApJ...475...20L, 1999AJ....117.2024F}.  We
will thus delay the study of Eq.~\eqref{eq:5} till
Sect.~\ref{sec:more-accurate-shear}, hoping also that a detailed
analysis of the unbalanced estimator will make the complex discussion
of the balanced one easier to follow and results interpreted.

An estimate of the shear can be used to obtain an estimate for the
projected, dimensionless mass distribution $\kappa(\vec\theta)$ of the
lens using well-know techniques \citep[see,
e.g.,][]{1996A&A...305..383S, 1998A&A...335....1L,
  1999A&A...348...38L}.  In general, because of the linear
relationship between $\gamma$ and $\kappa$, we can write
\begin{equation}
  \label{eq:6}
  \kappa(\vec\theta) = \bar\kappa + \int_\Omega
  \mathcal{D}_i(\vec\theta, \vec\theta') \gamma_i(\vec\theta') \,
  \diff^2 \theta' \; ,
\end{equation}
where the Einstein's convention on repeated indeces has been used.  In
this equation, $\bar\kappa$ is the mean value of $\kappa$ on $\Omega$
and $\mathcal{D}_i(\vec\theta, \vec\theta')$ is a suitable kernel.
For example, in the limit where the area of observation $\Omega$ is
much larger than the size of the lensing cluster, a simple kernel can
be used \citep{KS}:
\begin{equation}
  \label{eq:7}
  \begin{pmatrix}
    \mathcal{D}^\mathrm{KS}_1 \\
    \mathcal{D}^\mathrm{KS}_2    
  \end{pmatrix}(\vec\theta, \vec\theta + \vec\phi) = \frac{1}{\pi
  \lvert \vec\phi \rvert^4}
  \begin{pmatrix}
    \phi_1^2 - \phi_2^2 \\
    2 \phi_1 \phi_2
  \end{pmatrix} \; .
\end{equation}
Note that in this case Eq.~\eqref{eq:6} is a simple convolution.  For
finite fields $\Omega$, the kernel $\mathcal{D}_i(\vec\theta,
\vec\theta')$ often cannot be written in closed form, still it can be
written in terms of the Green function for a Neumann problem
\citep[see][]{1998A&A...335....1L, 2001A&A...374..740S}.  Fast
numerical methods exist to obtain $\kappa$ from the shear $\gamma$
\citep{1999A&A...348...38L}.

The linear relationship between $\kappa$ and $\gamma$ allows us to
easily derive the statistical properties of the mass map from the
properties of the shear.  In the following, thus, we will first
evaluate the expectation value and the noise of the shear, and then
obtain analogous results for the mass map.

\section{Statistical analysis}
\label{sec:statistical-analysis}

In this section we will perform a statistical analysis of the shear
estimator \eqref{eq:4} using a general probability density
distribution for all random variables involved.  As already noted
above, this estimator is not optimal and generally should not be used.
One purpose of this paper is actually to show that the increase in
noise of unbalanced estimator is very large in most cases.

Let us describe in detail the various random variables used.
\begin{description}
\item[\bf Source ellipticities $\epsilon^{\mathrm{s}(n)}$.] We assume
  that the intrinsic ellipticity of each galaxy follows the
  probability distribution $p_\epsilon(\epsilon^\mathrm{s})$; this
  probability distribution is taken to be isotropic, so that $\langle
  \epsilon^{\mathrm{s}(n)} \rangle = 0$.  Moreover, we assume no
  correlation between ellipticities of different galaxies, and no
  correlation between ellipticities and other random variables.
\item[\bf Source redshifts $z^{(n)}$.] The redshifts of sources are
  taken to be \textit{unknown\/} in the first part of this paper (see
  Sect.~\ref{sec:case-source-with} for the case where some knowledge
  on redshifts is available).  We assume that each redshift follows
  the probability distribution $p_z(z)$.  Correlations in redshifts and
  between redshifts and positions are considered.
\item[\bf Observed position $\vec\theta^{(n)}$.] The positions of
  galaxies are obviously readily available in weak lensing
  observations.  However, in order to obtain general results, we will
  perform an \textit{ensemble averaging\/} on galaxy positions.  This
  way, we will obtain results which are not strictly dependent on the
  particular configuration of galaxies.  Here we will assume that
  observed positions follow a uniform distribution on the field of
  observation $\Omega$ (thus neglecting the magnification bias).
  Correlations in positions and between positions and redshifts are
  considered.
\end{description}
In summary, the average of a function $f$ of random variables of a
single galaxy will be performed using the relation
\begin{align}
  \label{eq:8}
  \Bigl\langle f & \bigl(\epsilon^{(n)}, z^{(n)}, \vec\theta^{(n)}
  \bigr) \Bigr\rangle = \frac{1}{A} \int_\Omega \! \diff^2 \theta
  \int_0^\infty \!\!\!\! p_z(z) \, \diff z \notag\\
  & {} \times \int_{\lvert \epsilon^\mathrm{s} \rvert < 1}
  \hspace{-1.5em} p_\epsilon(\epsilon^\mathrm{s}) f\bigl(
  \epsilon^\mathrm{s} + Z(z) \gamma(\vec\theta), z, \vec\theta \bigr)
  \, \diff^2 \epsilon^\mathrm{s} \; .
\end{align}
Here $A$ is the area of the observation set $\Omega$.  Similarly, the
average of the product of two functions $f$ and $g$ of random
variables of two \textit{different\/} galaxies ($n \neq m$) will be
performed using the relation
\begin{align}
  \label{eq:9}
  \Bigl\langle f & \bigl(\epsilon^{(n)}, z^{(n)}, \vec\theta^{(n)}
  \bigr) g \bigl(\epsilon^{(m)}, z^{(m)}, \vec\theta^{(m)} \bigr)
  \Bigr\rangle = \notag\\
  & \Bigl\langle f \bigl(\epsilon^{(n)}, z^{(n)}, \vec\theta^{(n)}
  \bigr) \Bigr\rangle \cdot \Bigl\langle g \bigl(\epsilon^{(m)},
  z^{(m)}, \vec\theta^{(m)} \bigr) \Bigr\rangle \notag\\
  & {} + \frac{1}{A^2} \int_\Omega \! \diff^2 \theta \int_\Omega \!
  \diff^2 \theta' \int_0^\infty \!\!\!\! p_z(z) \, \diff z
  \int_0^\infty \!\!\!\! p_z(z') \, \diff z' \notag\\
  & {} \times \int_{\lvert \epsilon^\mathrm{s} \rvert < 1}
  \hspace{-1.5em} p_\epsilon(\epsilon^\mathrm{s}) \diff^2
  \epsilon^\mathrm{s} \int_{\lvert \epsilon^{\mathrm{s}\prime} \rvert
    < 1} \hspace{-1.5em} p_\epsilon(\epsilon^{\mathrm{s}\prime})
  f(\epsilon, z, \vec\theta) g(\epsilon', z', \vec\theta') \xi \,
  \diff^2 \epsilon^{\mathrm{s}\prime} \; .
\end{align}
In this relation, for simplicity, we have kept the original arguments
$\epsilon = \epsilon^\mathrm{s} + Z(z) \gamma(\vec\theta)$ and
$\epsilon' = \epsilon^{\mathrm{s}\prime} + Z(z') \gamma(\vec\theta')$
for $f$ and $g$ in the last term.  The function $\xi$, for our
purposes, can be written as
\begin{equation}
  \label{eq:10}
  \xi = \xi(\vec\theta - \vec\theta', z, z') \; .
\end{equation}
The galaxy two-point correlation function is often written in terms of
the proper distance $d$ between the galaxies.  Observations show that
the \textit{local\/} galaxy two-point correlation function is very
well approximated in the range $10 \mbox{ kpc} < hd < 10 \mbox{ Mpc}$
by the power law
\begin{equation}
  \label{eq:11}
  \xi(d) = \left( \frac{d_0}{d} \right)^\eta \; ,
\end{equation}
with $d_0 \simeq (5.4 \pm 1) \, h^{-1} \mbox{ Mpc}$, and $\eta
\simeq 1.77 \pm 0.04$ (Peebles 1993).  At high redshift ($z \sim 1$),
the two-point correlation function is more poorly known.  The
redshift-dependence is often written as a power-law  
\begin{equation}
  \label{eq:12}
  \xi(d, z) = \left( \frac{d_0}{d} \right)^\eta (1 + z)^\alpha \; .
\end{equation}
Recent investigations \citep{1996ApJ...461..534L} suggests that
$\alpha$ is in the range $[-3, -1]$.  In this paper, then, we will
take the fiducial value $\alpha = -2$.

Finally, we note that, given the linear relationship between
$\epsilon$ and $\epsilon^\mathrm{s}$, it is not difficult to include
the effect of errors of measurements for the observed ellipticities in
our discussion.  In general, such errors can be described using a
Baysian probability distribution $p_\mathrm{obs}(\hat\epsilon |
\epsilon)$ which gives the probability of measuring $\hat\epsilon$
when the real lensed ellipticity is $\epsilon$.  Then the standard
framework described above can be applied without any change if we
replace the probability distribution $p(\epsilon^\mathrm{s})$ with an
\textit{effective\/} distribution
$p_\mathrm{eff}(\epsilon^\mathrm{s})$ given by
\begin{equation}
  \label{eq:13}
  p_\mathrm{eff}(\hat\epsilon^\mathrm{s}) = \int_{\lvert
  \epsilon^\mathrm{s} \rvert < 1}
  p_\mathrm{obs}(\hat\epsilon^\mathrm{s} | \epsilon^\mathrm{s})
  p_\epsilon(\epsilon^\mathrm{s}) \, \diff^2 \epsilon^\mathrm{s} \; .
\end{equation}
We stress that, because of the linearity of Eq.~\eqref{eq:1}, the
distribution $p_\mathrm{obs}(\hat\epsilon^\mathrm{s} |
\epsilon^\mathrm{s})$ is formally identical to the distribution
$p_\mathrm{obs}(\hat\epsilon | \epsilon)$.  Finally, we observe that if
this Baysian distribution is taken to depend only on the difference
$\lvert \hat\epsilon - \epsilon \rvert$, then the \textit{effective
  variance\/} $\sigma_{\epsilon\mathrm{eff}}^2$ of
$\epsilon^\mathrm{s}$ can be written as
\begin{equation}
  \label{eq:14}
  \sigma_{\epsilon\mathrm{eff}}^2 = \sigma_\epsilon^2 +
  \sigma_\mathrm{obs}^2 \; ,
\end{equation}
where $\sigma_\mathrm{obs}^2$ is the variance of $(\hat\epsilon -
\epsilon)$ (given by $p_\mathrm{obs}$), and $\sigma^2_\epsilon$ is the
variance of the source ellipticities, defined as $\bigl\langle
\epsilon^\mathrm{s}_i \epsilon^\mathrm{s}_j \bigr\rangle =
\sigma^2_\epsilon \delta_{ij}$.

\subsection{Average value}
\label{sec:average-value}

The average value of the estimator \eqref{eq:4} is easily evaluated
using Eq.~\eqref{eq:8}:
\begin{equation}
  \label{eq:15}
  \bigl\langle \hat\gamma(\vec\theta) \bigr\rangle = \int_\Omega
  W(\vec\theta - \vec\theta') \gamma(\vec\theta') \, \diff^2
  \theta' \; .
\end{equation}
Because of the linearity between $\gamma$ and $\kappa$ we also obtain
\begin{equation}
  \label{eq:16}
  \bigl\langle \hat\kappa(\vec\theta) \bigr\rangle = \int_\Omega
  W(\vec\theta - \vec\theta') \kappa(\vec\theta') \, \diff^2
  \theta' \; .  
\end{equation}
In other words, the expected value of the estimated mass map is the
true mass map smoothed by the weight function $W$.  The length-scale
of this function, thus, sets the resolution of the shear and mass map.

\subsection{Covariance}
\label{sec:covariance}

A complete characterization of the error on the shear map is given by
the \textit{two-point correlation function}, defined as
\citep[see][]{1998A&A...335....1L}
\begin{equation}
  \label{eq:17}
  \Cov_{ij}(\hat\gamma; \vec\theta, \vec\theta') = \Bigl\langle 
  \bigl[ \hat\gamma_i(\vec\theta) - \bigl\langle
  \hat\gamma_i(\vec\theta) \bigr\rangle \bigr] 
  \bigl[ \hat\gamma_j(\vec\theta') - \bigl\langle
  \hat\gamma_j(\vec\theta') \bigr\rangle \bigr] \Bigr\rangle \; .
\end{equation}
Analogously, we define the two-point correlation function for the mass
map as
\begin{equation}
  \label{eq:18}
  \Cov(\hat\kappa; \vec\theta, \vec\theta') = \Bigl\langle 
  \bigl[ \hat\kappa(\vec\theta) - \bigl\langle \hat\kappa(\vec\theta)
  \bigr\rangle \bigr]
  \bigl[ \hat\kappa(\vec\theta') - \bigl\langle \hat\kappa(\vec\theta')
  \bigr\rangle \bigr] \Bigr\rangle \; .
\end{equation}
Equation~\eqref{eq:6} provides a simple way to obtain
$\Cov(\hat\kappa)$ from $\Cov(\hat\gamma)$:
\begin{align}
  \label{eq:19}
  \Cov(\hat\kappa; \vec\theta, \vec\theta') = {} & \int_\Omega \diff^2
  \phi \, \mathcal{D}_i(\vec\theta, \vec\phi) \int_\Omega \diff^2
  \phi' \, \mathcal{D}_j(\vec\theta', \vec\phi') \notag\\
  & {} \times \Cov_{ij}(\hat\gamma; \vec\phi, \vec\phi') \; .
\end{align}
We stress that the simple variance of $\hat\gamma$, i.e.\ the quantity
$\Cov_{ii}(\hat\gamma, \vec\theta, \vec\theta)$, would not be
sufficient to evaluate the variance of $\hat\kappa$.

In order to obtain $\Cov(\hat\gamma)$ we write the scatter of $\hat
\gamma$ from its average value as
\begin{align}
  \label{eq:20}
  \bigl[\hat\gamma_i - \langle \hat\gamma_i \rangle \bigr](\vec\theta)
  = {} & \frac{1}{\langle Z \rangle \rho} \sum_{n=1}^N \biggl[ \Bigl(
  \epsilon_i^{\mathrm{s}(n)} + \gamma_i \bigl( \vec\theta^{(n)} \bigr)
  Z^{(n)} \Bigr) W\bigl( \vec\theta - \vec\theta^{(n)} \bigr) \notag\\
  & {} - \frac{\langle Z \rangle}{A} \int_\Omega W(\vec\theta -
  \vec\phi) \gamma_i(\vec\phi) \, \diff^2 \phi \biggr] \; ,
\end{align}
since $N = A \rho$.  Hence we find
\begin{align}
  \label{eq:21}
  \Cov_{ij}(\hat\gamma; \vec\theta, \vec\theta') & {} 
  = {} \notag\\
  \frac{1}{\langle Z \rangle^2 \rho^2} \Bigg\langle \sum_{n=1}^N
  \biggl[ & \Bigl( \epsilon_i^{\mathrm{s}(n)} + \gamma_i \bigl(
  \vec\theta^{(n)} \bigr) Z^{(n)} \Bigr) W\bigl( \vec\theta -
  \vec\theta^{(n)} \bigr) \notag\\
  & {} - \frac{\langle Z \rangle}{A} \int_\Omega W(\vec\theta -
  \vec\phi) \gamma_i(\vec\phi) \, \diff^2 \phi \biggr] \notag\\
  {} \times \sum_{m=1}^N \biggl[ &
  \Bigl( \epsilon_j^{\mathrm{s}(m)} + \gamma_j \bigl( \vec\theta^{(m)}
  \bigr) Z^{(m)} \Bigr) W\bigl( \vec\theta' - \vec\theta^{(m)}
  \bigr) \notag\\
  & {} - \frac{\langle Z \rangle}{A} \int_\Omega W(\vec\theta' -
  \vec\phi') \gamma_j(\vec\phi') \, \diff^2 \phi' \biggr]
  \Biggr\rangle\; .
\end{align}
This expression is evaluated more easily if we distinguish the two
situations $n = m$ and $n \neq m$.  Let us start with the case $n =
m$.

If $n = m$, then we are considering the same galaxy and
Eq.~\eqref{eq:8} must be applied.  In this case, using the isotropy
hypothesis for source ellipticities, we obtain
\begin{align}
  \label{eq:22}
  C_1 = {} & \frac{\sigma^2_\epsilon \delta_{ij}}{\langle Z \rangle^2
    \rho} \int_\Omega W(\vec\theta - \vec\phi) W(\vec\theta' -
  \vec\phi) \, \diff^2 \phi \notag\\
  & {} + \frac{\bigl\langle Z^2 \bigr\rangle}{\langle Z \rangle^2
    \rho} \int_\Omega \gamma_i(\vec\phi) \gamma_j(\vec\phi)
  W(\vec\theta
  - \vec\phi) W(\vec\theta' - \vec\phi) \, \diff^2 \phi \notag\\
  & {} - \frac{1}{\rho^2 A^2} \int_\Omega W(\vec\theta - \vec\phi)
  \gamma_i(\vec\phi) \, \diff^2 \phi
  \notag\\
  & \phantom{{} - \frac{1}{\rho^2 A}} {} \times \int_\Omega
  W(\vec\theta' - \vec\phi') \gamma_j(\vec\phi') \, \diff^2 \phi' \; .
\end{align}
The last term vanishes in the limit where the area of observation $A$
is large.  [What actually matters here is the ratio between the
``effective'' area of the weight function $W$ and $A$.]\@ We thus will
ignore this term and similar terms arising in other expressions.

When $n \neq m$, on the other hand, we obtain a single term involving
the two-point correlation function
\begin{align}
  \label{eq:23}
  C_2 = {} & \frac{1}{\langle Z \rangle^2} \int_\Omega \! \diff^2 \phi
  W(\vec\theta - \vec\phi) \gamma_i(\vec\phi) \int_\Omega \!  \diff^2
  \phi' W(\vec\theta' - \vec\phi') \gamma_j(\vec\phi') \notag\\
  & {} \times \int_0^\infty \!\!\!\! p_z(z) Z(z) \, \diff z
  \int_0^\infty \!\!\!\! p_z(z') Z(z') \xi(\vec\phi -
  \vec\phi', z, z') \, \diff z' \; .
\end{align}
The last two integrals in the r.h.s.\ of this equation are reminiscent
of Limber's equation \citep[see][]{Peebles} and could suggest that the
(projected) angular correlation function is relevant for lensing.
This function, in fact, can be written in term of the full
three-dimensional correlation $\xi$ as
\begin{equation}
  \label{eq:24}
  w(\vec\varphi) = \int_0^\infty \!\!\!\! p_z(z) \, \diff z
  \int_0^\infty \!\!\!\! p_z(z') \xi(\vec\varphi, z, z') \, \diff z' \; .
\end{equation}
In reality, because of the presence of the cosmological weights $Z(z)$
and $Z(z')$ in the integrand of Eq.~\eqref{eq:23}, the expression for
$C_2$ cannot be rewritten in terms of the angular correlation
$w(\vec\varphi)$ only.

To simplify the notation, we will write terms similar to the one found
in Eq.~\eqref{eq:23} as
\begin{equation}
  \label{eq:25}
  C_2 = \frac{1}{\langle Z \rangle^2} \Xi\bigl[ W \gamma_i Z  W'
  \gamma_j' Z' \bigr] \; .
\end{equation}
The functional $\Xi$ represents a multiple integration over
$\xi(\vec\phi - \vec\phi', z, z') \, \diff^2 \phi \, \diff^2 \phi' \,
\diff z \, \diff z'$.  In summary, the correlation for the shear can
be written as
\begin{align}
  \label{eq:26}
  \Cov_{ij} & (\hat\gamma; \vec\theta, \vec\theta') =
  \frac{\sigma^2_\epsilon \delta_{ij}}{\langle Z \rangle^2 \rho}
  \int_\Omega W(\vec\theta -
  \vec\phi) W(\vec\theta' - \vec\phi) \, \diff^2 \phi \notag\\
  &{} + \frac{\bigl\langle Z^2 \bigr\rangle}{\langle Z \rangle^2 \rho}
  \int_\Omega \gamma_i(\vec\phi) \gamma_j(\vec\phi) W(\vec\theta -
  \vec\phi) W(\vec\theta' - \vec\phi) \,
  \diff^2 \phi \notag\\
  &{} + \frac{1}{\langle Z \rangle^2} \Xi\bigl[ W \gamma_i Z W'
  \gamma_j' Z' \bigr] \; .
\end{align}
This expression, although rather complicated, already clearly shows
the various contribution to the noise.  The first term on the r.h.s.,
proportional to $\sigma_\epsilon^2$, is due to intrinsic scatter of
source ellipticities.  \citet{1998A&A...335....1L} have already
considered this term, and have also shown that because of its
``diagonal'' form (it is proportional to $\delta_{ij}$ and is a simple
convolution of $W$ with itself) can be easily converted into a
covariance for $\kappa$.  Writing $\bigl\langle Z^2 \bigr\rangle =
\langle Z \rangle^2 + \sigma_Z^2$, the second term on the r.h.s.\ can
be thought as the sum of two terms.  One term does not involve any
redshift dependence and is a Poisson noise on the angular density of
galaxies.  The other term is proportional to $\sigma^2_Z$, the scatter
of galaxies in redshift, and is a Poisson noise on the redshift
distribution.  These two terms can be better understood by two simple
examples.  Suppose that we observe some gravitational lens and that we
make a shear map using the estimator of Eq.~\eqref{eq:4}.  Because of
Poisson noise, in some region of the sky we could observe an
overdensity of background galaxies, and thus obtain a larger than
expected value for the shear in that region.  On the other hand, in
another region we could observe a ``normal'' density of galaxies, but
their redshifts, because of Poisson noise, could be larger than
expected.  As a result, in this second region too we would
over-estimate the shear.  Situations such as the ones described
contribute to the noise on $\hat\gamma$ by adding two sources of
errors characterized by their dependence on $\gamma$ (in both cases
considered above, in fact, no error is expected if $\gamma = 0$).
Finally, turning to the last term of Eq.~\eqref{eq:26}, we note that
it is clearly related to the two-point galaxy correlation function.
Analogously to the second term of Eq.~\eqref{eq:26}, this is actually
composed of two contribution, one due to the angular correlation and
one due to correlation in redshift [cf.\ Eqs.~\eqref{eq:23} and
\eqref{eq:24}].

From Eq.~\eqref{eq:26} we can also see that the first two noise terms
are proportional to $1/\rho$, while the last one is
\textit{apparently\/} independent of the density of background
galaxies.  This behavior is easily explained.  Ellipticity and Poisson
variances depends on single galaxies and thus increase linearly with
$\rho$; hence, the factor $1/\rho$ in Eq.~\eqref{eq:4} makes the final
noise due to these sources proportional to $1/\sqrt{\rho}$.  On the
other hand, the variance due to galaxy correlation is proportional to
the number of \textit{pairs\/} of galaxies and thus to $\rho^2$; the
final correlation noise is then formally independent of $\rho$.  We
stress, however, that all noise terms depend on the galaxy redshift
distribution.  The density $\rho$ and the probability distribution
$p_z(z)$ are intimately related.  For example, deeper observations
will increase $\rho$ but also the mean galaxy redshift.  As a result,
all noise terms \textit{do depend\/} on the depth of the observation
(see Sect.~\ref{sec:impact-observations} for further comments on this
point).

Looking again at Eq.~\eqref{eq:26}, we can also obtain information
about the typical \textit{correlation length\/} for the shear, i.e.\ 
the largest angular separation between points that have correlated
noise properties.  The first term in this expression, in fact,
suggests that the correlation length for $\hat \gamma$ is
approximately twice the length scale of the weight function $W$; the
same is true for the second term, while the third term has no
``intrinsic'' scale.  This last point is basically due to the lack of
a scale of the two-point correlation function as given by
Eq.~\eqref{eq:11} (note that $d_0$ in this expression should be taken
as a simple multiplicative coefficient rather than as a scale for
$\xi$).

In principle, for a given lens, we could now evaluate the covariance
of $\hat\gamma$ using the results obtained so far, and then convert it
to the covariance of $\kappa$.  Actually, in the limit of large fields
$\Omega$, we can use the kernel \eqref{eq:7} in Eq.~\eqref{eq:19},
thus obtaining directly an expression for the covariance of $\kappa$:
\begin{align}
  \label{eq:27}
  \Cov & (\hat\kappa; \vec\theta, \vec\theta') =
  \frac{\sigma^2_\epsilon}{\langle Z \rangle^2 \rho} \int_\Omega
  \tilde W_i(\vec\theta - \vec\phi) \tilde W_i(\vec\theta' - \vec\phi)
  \, \diff^2 \phi \notag\\ 
  &{} + \frac{\bigl\langle Z^2 \bigr\rangle}{\langle Z \rangle^2 \rho}
  \int_\Omega \gamma_i(\vec\phi) \gamma_j(\vec\phi)
  \tilde W_i(\vec\theta - \vec\phi) \tilde W_j(\vec\theta' - \vec\phi) \,
  \diff^2 \phi \notag\\ 
  &{} + \frac{1}{\langle Z \rangle^2} \Xi\bigl[ \tilde W_i \gamma_i Z
  \tilde W_j' \gamma_j' Z' \bigr] \; .
\end{align}
Here, $\tilde W_i$ is a ``modified weight,'' i.e.\ the convolution of
$W$ with $\mathcal{D}^\mathrm{KS}_i$: 
\begin{align}
  \label{eq:28}
  \tilde W_i(\vec\theta) & {} = \int W(\vec\theta - \vec\phi)
  \mathcal{D}^\mathrm{KS}_i(\vec\phi) \, \diff^2 \phi \notag\\
  & {} = \mathcal{D}^\mathrm{KS}_i(\vec\theta) \left[ 1 - \left( 1 +
      \frac{\lvert \vec\theta \rvert^2}{2 \sigma_W^2} \right) \exp
    \left( - \frac{\lvert \vec\theta \rvert^2}{2 \sigma_W^2} \right)
  \right]\; .
\end{align}
The last equality holds when a Gaussian weight function of the form
\eqref{eq:3} is used.

\section{Slowly varying shear}
\label{sec:slowly-varying-fields}

The results obtained in the last section are exact but, unfortunately,
difficult to interpret in general.  In particular, it is \textit{not
  easy\/} to estimate the order of magnitude of the expressions
obtained, and it is also difficult to say which term makes the more
important contribution in Eq.~\eqref{eq:26} [or, equivalently,
Eq.~\eqref{eq:27}].  In an important limit, however, the results
obtained so far can be written in a much simpler and more useful form.

Suppose that the true shear field $\gamma(\vec\theta)$ is slowly
varying on the typical scale of the weight function $W$.  In this
limit, we can expand the shear to first-order and greatly simplify the
integrals that involve $\gamma$ and $W$.  The results obtained, though
clearly not exact, represent an extremely useful starting point to
evaluate the noise of the proposed shear estimates.  Note that this
technique can be successfully used to simplify the expressions of
statistical estimates for the shear field, while is not directly
applicable to expressions for the mass map because of the non-local
dependence of $\kappa$ on $\gamma$ (or, equivalently, because of the
infinite support of $\mathcal{D}^\mathrm{KS}$).  Actually, we will
still be able to obtain a finite form for the average value of $\hat
\kappa(\vec\theta)$ (see Sect.~\ref{sec:average-value}), but not for
the covariance.  However, if the noise properties of the shear are
known, the covariance of $\hat \kappa$ can be evaluated using
Eq.~\eqref{eq:19}.  Moreover, it is not unreasonable to assume that
the covariance of $\hat\kappa$ is close to $\Cov(\hat\gamma)$
[actually, if only the first term of Eq.~\eqref{eq:26} is taken into
account, the two covariances are identical].

In the following we will \textit{always\/} assume a symmetric weight
function, i.e.\ such that $W(\vec\theta - \vec\theta')$ depends only
on $\lvert \vec\theta - \vec\theta' \rvert$.  Moreover, the results
obtained will also be specialized to the important case where $W$ is a
Gaussian of the form \eqref{eq:3}.  Finally, in order to numerically
estimate the contribution of some terms to the noise, we will often
refer to a ``typical'' model.  The model is characterized by a lens at
redshift $z_\mathrm{d} = 0.3$ in an Einstein-de~Sitter cosmological
model.  The normalized redshift probability distribution of source
galaxies will be taken to be \citep[see][]{1996ApJ...466..623B}
\begin{equation}
  \label{eq:29}
  p_z(z) = \frac{\beta z^2}{\Gamma(3/\beta) z_0^3} \exp\left[-\left(
  \frac{z}{z_0} \right)^\beta \right] \; .
\end{equation}
The constant $z_0$ determines the depth of the observations, while
$\beta$ determines the steepness of the cutoff at high redshifts.  The
moments of this distribution are
\begin{equation}
  \label{eq:30}
  \bigl\langle z^k \bigr\rangle = z_0^k \frac{\Gamma\bigl( (3 +
  k)/\beta \bigr)}{\Gamma(3/\beta)} \; ,
\end{equation}
and the mode is
\begin{equation}
  \label{eq:31}
  z_\mathrm{mode} = z_0 \left( \frac{2}{\beta} \right)^{1/\beta}
  \; .
\end{equation}
In the following, when not otherwise stated, we will use $\beta = 1.5$
and $z_0 = 0.7$, so that $\langle z \rangle \simeq 1.05$ and
$z_\mathrm{mode} \simeq 0.848$.

\subsection{Average value}
\label{sec:average-value-3}

If $\gamma(\vec\theta')$ does not change significantly when
$\vec\theta'$ is on the effective area of $W$, we can expand the shear
around the point $\vec\theta$:
\begin{equation}
  \label{eq:32}
  \gamma_i(\vec\theta') \simeq \gamma(\vec\theta) + \left. \frac{\partial
  \gamma_i}{\partial \theta_j} \right|_{\vec\theta} (\theta'_j - \theta_j)
  \; .
\end{equation}
Inserting this expression into Eq.~\eqref{eq:15}, and using the symmetry
of $W$, we get
\begin{equation}
  \label{eq:33}
  \bigl\langle \hat \gamma(\vec\theta) \bigr\rangle =
  \gamma(\vec\theta) \int_\Omega W(\vec\theta - \vec\theta') \,
  \diff^2 \theta' = \gamma(\vec\theta) \; .
\end{equation}

Equation~\eqref{eq:33} is clearly a rough approximation, since we
already know that average value for the measured shear map is a
\textit{smoothed\/} version of the true shear.  On the other hand, if
the shear does not change significantly on the scale of $W$, we can
safely take the result obtained; in more general cases it is, at
least, an order-of-magnitude estimate.

\subsection{Covariance ($\xi = 0$)}
\label{sec:covariance-xi-0}

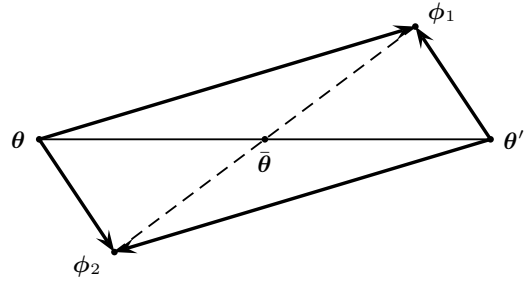
\begin{figure}[t!]
  \begin{center}
    \psset{unit=1cm}
    \begin{pspicture}(-3.5,-2)(3.5,2)
      \psdots(-3,0)(3,0)(-2,-1.5)(2,1.5)(0,0)
      \psset{linewidth=0.7pt}
      \psline(-3,0)(3,0)
      \psline[linestyle=dashed](-2,-1.5)(2,1.5)
      \psset{linewidth=1.3pt}
      \psline{->}(-3,0)(2,1.5)
      \psline{->}(3,0)(2,1.5)
      \psline{->}(-3,0)(-2,-1.5)
      \psline{->}(3,0)(-2,-1.5)
      \uput[180](-3,0){$\vec\theta$}
      \uput[0](3,0){$\vec\theta'$}
      \uput[30](2,1.5){$\vec\phi_1$}
      \uput[210](-2,-1.5){$\vec\phi_2$}
      \uput[-90](0,0){$\bar{\vec\theta}$}
    \end{pspicture}
    \caption{Linear terms of the expansion of $\gamma$ around
      $\bar{\vec\theta}$ do not enter the expression for the
      covariance.  In fact, the integrand of any linear term evaluated
      at the point $\vec\phi_1$ is the opposite of the integrand
      evaluated at $\vec\phi_2$ ($\vec\phi_1$ and $\vec\phi_2$ are
      such that $\vec\phi_1 + \vec\phi_2 = 2 \bar{\vec\theta}$).
      In particular, the application of this symmetry to the liner
      expansion for the term $\gamma_i(\vec\phi) \gamma_j(\vec\phi)$
      leads to Eq.~\eqref{eq:34}.}
    \label{fig:1}
  \end{center}
\end{figure}

We can in a similar way obtain an expression for the covariance of the
shear using Eq.~\eqref{eq:26}.  The method used is similar to the one
adopted above, with the only small difference that now the expansion
for $\gamma$ will be performed around the mid-point $\vecbar\theta =
(\vec\theta + \vec\theta') / 2$ (see Fig.~\ref{fig:1}).  We thus find
\begin{align}
  \label{eq:34}
  &\Cov_{ij}(\hat\gamma; \vec\theta, \vec\theta') = \frac{1}{\langle Z
    \rangle^2 \rho} \left[ \sigma^2_\epsilon \delta_{ij} +
    \bigl\langle Z^2\bigr\rangle \gamma_i(\vecbar\theta)
    \gamma_j(\vecbar\theta) \right] \notag\\
  & \qquad {} \times (W \star W)(\vec\theta - \vec\theta') \notag\\
  & \quad {} + \frac{\bigl\langle Z^2 \bigr\rangle}{\langle Z
    \rangle^2 \rho} \gamma_{i,i'} (\bar{\vec\theta}) \gamma_{j,j'}
  (\bar{\vec\theta})
  \notag\\
  & \qquad {} \times \int_\Omega W(\vec\theta - \vec\phi)
  W(\vec\theta' - \vec\phi) (\phi_{i'} - \bar\theta_{i'}) (\phi_{j'} -
  \bar\theta_{j'}) \, \diff^2 \phi \; ,
\end{align}
where the star ($\star$) represents a convolution.  In the case where
$W$ is a Gaussian function we obtain in particular
\begin{align}
  \label{eq:35}
  \Cov_{ij}(\hat\gamma; \vec\theta, \vec\theta') = {} & \frac{1}{4 \pi
    \langle Z \rangle^2 \rho \sigma_W^2} \biggl[ \sigma^2_\epsilon
  \delta_{ij} + \bigl\langle Z^2 \bigr\rangle \gamma_i(\vecbar\theta)
  \gamma_j(\vecbar\theta) \notag\\
  & \quad {} + \frac{\bigl\langle Z^2 \bigr\rangle \sigma_W^2}{2} 
  \gamma_{i,k} (\bar{\vec\theta}) \gamma_{j,k} (\bar{\vec\theta})
  \biggr]
  \notag\\
  & {} \times \exp \left( - \frac{\lvert \vec\theta - \vec\theta'
      \rvert^2}{4 \sigma_W^2} \right) \; ,
\end{align}
This is the result for the covariance of $\hat \gamma$ in the slowly
varying shear approximation.  The fact that a finite expression has
been obtained allows us to study the behavior of the various
contributions to the noise.  Let us discuss in detail the result
obtained.

Equation~\eqref{eq:34} shows that the noise is proportional to a
convolution of $W$ with itself.  This suggests to define the
\textit{effective area\/} $A_W$ of the weight function, i.e.\ the area
where $W(\vec\phi)$ is \textit{significantly\/} different from zero
or, formally,
\begin{equation}
  \label{eq:36}
  A_W = \biggl[ \int \bigl[ W(\vec\phi) \bigr]^2 \, \diff^2 \phi
  \biggr]^{-1} \; .
\end{equation}
For the Gaussian weight function we get $A_W = 4 \pi \sigma_W^2$.  We
then note that the noise on $\hat \gamma$ given by Eq.~\eqref{eq:35}
is proportional to $1/N_W$, where $N_W = \rho A_W$ is the expected
number of galaxies inside the effective area of $W$.

The ratio between the contributions to the noise arising from the
shot-noise and the intrinsic scatter in ellipticities is of order
$\bigl\lvert \gamma(\vec\theta) \bigr\rvert^2 \bigl\langle Z^2
\bigr\rangle / \sigma_\epsilon^2$.  Since $\sigma_\epsilon \simeq
0.25$, and $\bigl\langle Z^2 \bigr\rangle \sim 1$, we see that the
shot-noise is a relevant term unless the lens is very weak.  More
specifically, let us consider the error made on the component $i$ of
the shear at $\vec\theta$.  We have from Eq.~\eqref{eq:35} (no sum on
$i$)
\begin{align}
  \label{eq:37}
  \Cov_{ii}(\hat\gamma; \vec\theta, \vec\theta) & {} = \frac{1}{4 \pi
    \langle Z \rangle^2 \rho \sigma_W^2} \biggl[ \sigma^2_\epsilon
  + \bigl\langle Z^2 \bigr\rangle \bigl[ \gamma_i(\vec\theta) \bigr]^2
  \notag\\
  & \phantom{{} = {}} \quad {} + \frac{\bigl\langle Z^2 \bigr\rangle
    \sigma_W^2}{2} \gamma_{i,k} (\vec\theta) \gamma_{i,k} (\vec\theta) \biggr]
    \notag\\
  &{} = \Sigma^2_\epsilon + \Sigma^2_\mathrm{P0} + \Sigma^2_\mathrm{P1} \; ,
\end{align}
where, for convenience, we have explicitely split the three
contributions to the noise considered here.  Using the lens model
described above, we have $\langle Z \rangle \simeq 0.600810$ and
$\sigma^2_Z \simeq 0.038347$.  Hence, the expected error on
$\hat\gamma$ due to the spread of source ellipticities is
\begin{align}
  \label{eq:38}
  \Sigma_\epsilon = {} & 0.016600 \, \biggl(
  \frac{\sigma_\epsilon}{0.25} \biggr) \biggl( \frac{\sigma_W}{1
    \mbox{ arcmin}} \biggr)^{-1} \notag\\
  & {} \times \biggl( \frac{\rho}{50 \mbox{ gal arcmin}^{-2}}
  \biggr)^{-1/2} \; .
\end{align}
The Poisson noise contributes with a term proportional to the shear, 
\begin{align}
  \label{eq:39}
  \Sigma_\mathrm{P0} = {} & 0.004196 \, \biggl( \frac{\gamma}{0.1}
  \biggr) \biggl( \frac{\sigma_W}{1 \mbox{ arcmin}} \biggr)^{-1}
  \notag\\
  & {} \times \biggl( \frac{\rho}{50 \mbox{ gal arcmin}^{-2}}
  \biggr)^{-1/2} \! ,
\end{align}
and with a term proportional to the derivatives of the shear,
\begin{align}
  \label{eq:40}
  \Sigma_\mathrm{P1} = 0.002967 \, \biggl( \frac{\gamma_{,j}}{0.1
  \mbox{ arcmin}^{-1}} \biggr) \biggl( \frac{\rho}{50 \mbox{ gal
  arcmin}^{-2}} \biggr)^{-1/2} \! .
\end{align}
Note that, surprisingly, this term does not depend on $\sigma_W$ (in
this case, the angular scaling is given by the gradient of the shear).
In conclusion, for a typical weak lens at redshift $z_\mathrm{d} =
0.3$ we expect similar contributions from the intrinsic scatter of
ellipticities and Poisson noise.

\subsection{Covariance ($\xi \neq 0$)}
\label{sec:covariance-xi-neq}

Analytical calculations for the contributions to the covariance from
the galaxy two-point correlation are definitely
\textit{non-trivial\/}.  For this reason, we report here only the
results of such calculations, and we refer to
Appendix~\ref{sec:slowly-vary-fields-1} for any details.  Note that the
calculations have been carried out only in the (important) case where
$W$ is a Gaussian of the form \eqref{eq:3}.  Moreover, sensible
approximations have been used.

As discussed above [see Eqs.~\eqref{eq:11} and \eqref{eq:12}], the
two-point correlation function is well approximated by a simple
power-law in the physical separation of galaxies.  Moreover,
observations suggests that the exponent $\eta$ is very close to
$2$.  This allows us to carry out calculations, obtaining [see
Eq.~\eqref{eq:23}]
\begin{align}
  \label{eq:41}
  C_2 = {}& \frac{\pi^{3/2} d_0^2}{2 \langle Z \rangle^2
    \sigma_W} \gamma_i(\vecbar\theta) \gamma_j(\vecbar\theta) \,
  I_0 \left( \frac{\lvert \vec\varphi \rvert^2}{8
      \sigma_W^2} \right) \exp \left( - \frac{\lvert \vec\varphi
      \rvert^2}{8 \sigma_W^2} \right) \notag\\
  & {} \times \int_{z_d}^\infty \frac{(1 + z)^\alpha \bigl[p_z(z) Z(z)
    \bigr]^2}{D_A(z) D_P'(z)} \, \diff z \; ,
\end{align}
where $I_0$ is the modified Bessel function of first kind,
$\vec\varphi = \vec\theta' - \vec\theta$, and $D_A(z)$ and $D_P(z)$
are, respectively, the angular diameter distance and the proper
distance of an object at redshift $z$ [note that in Eq.~\eqref{eq:41}
the derivative of $D_P(z)$ is involved].  Note that, in
Eq.~\eqref{eq:41}, the smoothing length $\sigma_W$ has to be expressed
in radians.  Moreover, we have used as usual the notation
$\bar{\vec\theta}$ for the mid-point between $\vec\theta$ and
$\vec\theta'$.

In the typical case discussed here of a lens at redshift $z_\mathrm{d}
= 0.3$ and using $d_0 = 5.4 \, h^{-1} \mbox{ Mpc}$ we have for the
expected \textit{error\/} due to the correlation of galaxies,
$\Sigma_2 = \sqrt{C_2(\vec\theta, \vec\theta)}$,
\begin{equation}
  \label{eq:42}
  \Sigma_2 = 0.031173 \, \biggl( \frac{\gamma}{0.1} \biggr)
  \biggl( \frac{\sigma_W}{1 \mbox{ arcmin}} \biggr)^{-1/2} \; .
\end{equation}
We must thus deduce that, at least for the case considered here, the
galaxy two-point correlation function represents a major source of
error in weak lensing estimates.  A more detailed discussion on the
relative contributions of the various sources of errors is delayed
until Sect.~\ref{sec:impact-observations}.

\section{The balanced shear estimator}
\label{sec:more-accurate-shear}

As already pointed out, the estimator \eqref{eq:4} is very simple and
traditionally has been the first one used \citep{KS}.  On the other
hand, this estimator has a rather large noise and in particular
suffers from Poisson noise, due to the fact that the normalization
factor is constant both on the density of galaxies ($1 / \rho$) and on
the redshift distribution of sources ($1 / \langle Z \rangle$).

In this section we will study in detail the estimator \eqref{eq:5}
similarly to what we have already done in
Sects.~\ref{sec:statistical-analysis} and
\ref{sec:slowly-varying-fields} for the estimator \eqref{eq:4}.
However, since now the estimator is \textit{non-linear\/} on the
observed positions of galaxies, an analytical evaluation of the
average values and errors for this estimator would be hopeless without
any approximations.  For this reason we will linearize
Eq.~\eqref{eq:5} around the average values of the random variables
involved.  This technique, often adopted in statistics, is briefly
described in Appendix~\ref{sec:linear-approximation-1}.  Moreover, to
simplify the discussion, we will mostly report here only results,
referring to Appendices~\ref{sec:deta-stat-analys} and
\ref{sec:slowly-vary-fields-1} for the derivations.

\subsection{Linear approximation}
\label{sec:linear-approximation}

Let us call $N$ and $D$ the numerator and denominator of
Eq.~\eqref{eq:5}, so that $\hat\gamma_i = N_i / D$.  Using the linear
approximation (see Appendix~\ref{sec:linear-approximation-1}), we can
write the expectation value of $\hat \gamma_i$ as
\begin{equation}
  \label{eq:43}
  \langle \hat\gamma_i \rangle \simeq \frac{\langle N_i \rangle}{\langle
    D \rangle} - \frac{1}{\langle D \rangle^2} \Cov(N_i D)
  + \frac{\langle N_i \rangle}{\langle D \rangle^3} \Cov(D D) \; .
\end{equation}
The last two terms on the r.h.s.\ represent second-order corrections
and will often be ignored.  The covariance of $\hat\gamma$ can be
written as $\Cov_{ij}(\hat\gamma; \vec\theta, \vec\theta')$, i.e.\ as
a $2 \times 2$ matrix which depends on two spatial coordinates
$\vec\theta$ and $\vec\theta'$ \citep[see][]{1998A&A...335....1L}.
Omitting the two arguments $\vec\theta$ and $\vec\theta'$, we can
write
\begin{align}
  \label{eq:44}
  & \Cov_{ij}(\hat\gamma) = \frac{1}{\langle D \rangle \langle D'
    \rangle} \Cov(N_i N'_j) - \frac{\langle N_i \rangle}{\langle D
    \rangle^2 \langle D' \rangle} \Cov(D N'_j) \notag\\
  & \qquad {} - \frac{\langle N'_j \rangle}{\langle D \rangle \langle
    D' \rangle^2} \Cov(N_i D') + \frac{\langle N_i \rangle \langle
    N'_j \rangle}{\langle D \rangle^2 \langle D' \rangle^2} \Cov(D D')
  \; .
\end{align}
Here we have used the prime to denote quantities which must be
evaluated at $\vec\theta'$ rather then at $\vec\theta$.

As a general rule, the linear approximation can be used when the
denominator $D$ is expected to have a small (relative) variance.  This
is true if the expected number of galaxies inside the effective area
of the smoothing function $W$ is much larger than $1$.

We also note that in the linear approximation, the expression for the
covariance for $\gamma$ depends on the two-point galaxy correlation
function only, and not on higher-order correlation functions.  This
would not be true without the linear approximation.

\subsection{Slowly varying shear}
\label{sec:slowly-vary-fields}

Similarly to what we have seen above for the unbalanced estimator
\eqref{eq:4}, the statistical properties of the balanced estimator
\eqref{eq:5} are better understood in the limit where the shear is
slowly changing on the effective area of the weight function.  In this
case, as shown in detail in Appendix~\ref{sec:slowly-vary-fields-1},
we can explicitely write the expectation value and the covariance of
$\hat \gamma$.

Regarding the expectation value of $\hat \gamma$, we obtain a result
similar to Eq.~\eqref{eq:33}, i.e. $\bigl\langle
\hat\gamma(\vec\theta) \bigr\rangle = \gamma(\vec\theta)$.  In other
words, in the slowly varying shear approximation, the estimator
\eqref{eq:5} is unbiased.  Note that in more general cases we still
recover Eq.~\eqref{eq:15}, i.e.\ the average value for the estimator
$\hat \gamma$ is the true shear $\gamma$ convolved with the weight
function\footnote{Actually, a rigorous calculation shows that the
  average value for $\hat\gamma$ is the true shear $\gamma$ convolved
  with a kernel slightly different from $W$.  The kernel quickly
  converges to the weight function when the number of effective
  galaxies is large \citep[see][]{2001A&A...373..359L}.}.  Finally,
since Eq.~\eqref{eq:15} holds, we can immediately deduce that
Eq.~\eqref{eq:16} holds as well.

Turning to covariances, we can finally see the advantages of using a
balanced estimator for the shear.  In fact, in the slowly varying
shear approximation and neglecting the two-point correlation
function, we obtain for the covariance of $\hat \gamma$, in the case
where $W$ is a Gaussian function,
\begin{align}
  \label{eq:45}
  & \Cov_{ij}(\hat\gamma; \vec\theta, \vec\theta') = \frac{1}{4 \pi
    \langle Z \rangle^2 \rho \sigma_W^2} \biggl[ \sigma^2_\epsilon
  \delta_{ij} + \sigma^2_Z \gamma_i(\vecbar\theta)
  \gamma_j(\vecbar\theta) \notag\\
  & \qquad {} + \biggl( \frac{\bigl\langle Z^2 \bigr\rangle
    \sigma_W^2}{2} \delta_{i'j'} - \frac{\langle Z \rangle^2}{4}
  \varphi_{i'} \varphi_{j'} \biggr) \gamma_{i,i'} (\bar{\vec\theta})
  \gamma_{j,j'}
  (\bar{\vec\theta})  \biggr] \notag\\
  & \quad {} \times \exp \left( - \frac{\lvert \vec\phi \rvert^2}{4
      \sigma_W^2} \right) \; ,
\end{align}
with $\sigma^2_Z = \bigl\langle Z^2 \bigr\rangle - \bigl\langle Z
\bigr\rangle^2$.  This expression should be compared with the
analogous equation obtained for the estimator \eqref{eq:4}, namely
Eq.~\eqref{eq:35}.  The basic difference is given by the factor
$\bigl\langle Z^2 \bigr\rangle$ which replaces $\sigma^2_Z$ of
Eq.~\eqref{eq:45} or, in other words, by the presence of an extra term
$\langle Z \rangle^2$ in Eq.~\eqref{eq:35}.  This extra term is
precisely due to Poisson noise on the angular coordinate $\vec\theta$;
analogously, the term $\sigma^2_Z$ left can be interpreted as Poisson
noise in redshift.  In order to better appreciate this point, it is
interesting to evaluate the expected error of the balanced estimator
in the lens model described above.  In this case we have $\langle Z
\rangle \simeq 0.600810$ and $\sigma^2_Z \simeq 0.038347$.  Hence, the
use of a balanced estimator reduces the ``constant'' component of
Poisson noise by about a factor $4$ (i.e.\ a factor $16$ in variance).
More explicitely, the contribution of the redshift Poisson noise is
given by
\begin{align}
  \label{eq:46}
  \Sigma_\mathrm{P0} = {} & 0.001300 \, \biggl( \frac{\gamma}{0.1}
  \biggr) \biggl( \frac{\sigma_W}{1 \mbox{ arcmin}} \biggr)^{-1} \notag\\
  & {} \times \biggl( \frac{\rho}{50 \mbox{ gal arcmin}^{-2}}
  \biggr)^{-1/2} \, .
\end{align}
This expression replaces \eqref{eq:39}.  Note that, in contrast, the
Poisson noise due to shear variations [represented by the second line
in Eq.~\eqref{eq:45}], is only partially reduced by the balanced
estimator.  In particular, if $\vec\varphi = \vec 0$, no improvement is
obtained and Eq.~\eqref{eq:40} can still be used to have an estimate
of this noise source.

If the two-point galaxy correlation is taken into account using the
model \eqref{eq:12}, we have a supplementary term in the covariance.
For the lens considered here, this term is of the order (see
Appendix~\ref{sec:slowly-vary-fields-1})
\begin{equation}
  \label{eq:47}
  \Sigma_2 = 0.007152 \, \biggl( \frac{\gamma}{0.1} \biggr) \biggl(
  \frac{\sigma_W}{1 \mbox{ arcmin}} \biggr)^{-1/2} \; .
\end{equation}
Comparing this result with Eq.~\eqref{eq:42}, we see that the use of
the balanced estimator has significantly reduced the noise due to the
clustering of galaxies.

\section{Numerical estimates}
\label{sec:numerical-estimates}

In the previous sections, we have obtained estimates for the various
contributions to the noise of the shear using several approximations,
namely the weak lensing limit, the slowly varying shear
(Sect.~\ref{sec:slowly-varying-fields}), and the linear expansion of
the estimators (Sect.~\ref{sec:linear-approximation}).  Among these
approximations, probably the most critical one is that of the slowly
varying shear.  Hence, in order to test the reliability of the results
obtained so far, we have numerically evaluated the noise on the shear
estimator \eqref{eq:4} in a specific case.  Note that, instead, it is
extremely difficult to test directly the results for the balanced
estimator without using the linear approximation described in
Sect.~\ref{sec:linear-approximation}.  In this case, in fact, we would
need to average over the whole galaxy positions $\bigl\{
\vec\theta^{(n)} \bigr\}$, using also a probability distribution $p
\bigl( \vec\theta^{(1)}, \vec\theta^{(2)}, \dots, \vec\theta^{(N)}
\bigr)$ which includes effects of $n$-point correlation (actually,
even $N$, the total number of galaxies, should be taken as a random
variable with Poisson distribution).  Clearly, such an analysis would
be prohibitive.

\begin{figure}[!t]
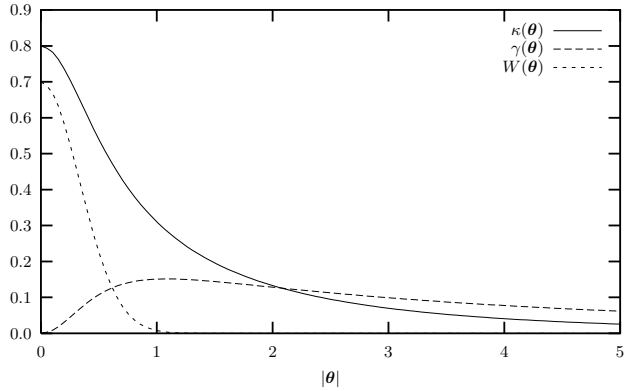

  \begin{center}
    \resizebox{\hsize}{!}{\input fig6.tex}
    \caption{The lens mass profile $\kappa(\vec\theta)$ and the lens
      shear $\gamma(\vec\theta)$ at different radii.  The plot also
      shows the Gaussian weight function $W(\vec\theta)$ used.  Notice
      that $w(\vec\theta)$ in this plot is not normalized.}
    \label{fig:2}
  \end{center}
\end{figure}

We have considered a lens at redshift $z_\mathrm{d} = 0.3$ with an
axisymmetric profile (see Fig.~\ref{fig:2}):
\begin{equation}
  \label{eq:48}
  \kappa(\vec\theta) = \frac{\kappa_0 \theta_\mathrm{t}}{%
    \theta_\mathrm{t} - \theta_\mathrm{c}} f\bigl( \lvert \vec\theta
  \rvert / \theta_\mathrm{c} \bigr) - \frac{\kappa_0 \theta_\mathrm{c}}{%
    \theta_\mathrm{t} - \theta_\mathrm{c}} f\bigl( \lvert \vec\theta
  \rvert / \theta_\mathrm{t} \bigr) \; ,
\end{equation}
where $f(x)$ is the dimensionless function
\begin{equation}
  \label{eq:49}
  f(x) = \frac{1}{\sqrt{1 + x^2}} \; .
\end{equation}
Note that $f(x) \propto 1/x$ for large $x$, $\kappa(\vec\theta)
\propto 1 / \lvert \vec\theta \rvert$ for $\theta_\mathrm{c} \ll
\lvert \theta \rvert \ll \theta_\mathrm{t}$, while $\kappa(\vec\theta)
\propto 1/\lvert \vec\theta \rvert^3$ for $\lvert \vec\theta \rvert
\gg \theta_\mathrm{t}$.  For this lens, the shear is easily obtained:
Its modulus is
\begin{equation}
  \label{eq:50}
  \gamma(\vec\theta) = \frac{\kappa_0 \theta_\mathrm{t}}{%
    \theta_\mathrm{t} - \theta_\mathrm{c}} g\bigl( \lvert \vec\theta
  \rvert / \theta_\mathrm{c} \bigr) - \frac{\kappa_0 \theta_\mathrm{c}}{%
    \theta_\mathrm{t} - \theta_\mathrm{c}} g\bigl( \lvert \vec\theta
  \rvert / \theta_\mathrm{t} \bigr) \; ,
\end{equation}
with
\begin{equation}
  \label{eq:51}
  g(x) = \frac{\bigl[ \sqrt{1 + x^2} - 1\bigr]^2}{x^2 \sqrt{1 + x^2}}
  \; .
\end{equation}
The shear decreases as $\gamma(\vec\theta) \propto 1/\lvert \vec\theta
\rvert^2$ for large $\lvert \vec\theta \rvert$ (see Fig.~\ref{fig:2}).

\begin{table}[!b]
  \centering
  \caption{Relevant parameters used for the numerical integration.}
  \label{tab:1}
  \small
  \begin{tabular}{ccl}
    Variable & Value & Comment \\
    \hline
    $z_\mathrm{d}$ & $0.3$ & Lens redshift\\
    $\theta_\mathrm{c}$ & $30''$ & Lens core radius
    [Eq.~\eqref{eq:48}]\\
    $\theta_\mathrm{t}$ & $5'$ & Lens truncation radius
    [Eq.~\eqref{eq:48}]\\
    $\kappa_0$ & $0.8$ & Lens central density [Eq.~\eqref{eq:48}]\\
    $\rho$ & $50 \mbox{ arcmin}^{-2}$ & Density of galaxies\\ 
    $\sigma_\epsilon$ & $0.25$ & Intrinsic ellipticity scatter\\
    $z_0$ & $0.7$ & Redshift distr.\ parameter [Eq.~\eqref{eq:29}]\\
    $\beta$ & $1.5$ & Redshift distr.\ slope [Eq.~\eqref{eq:29}]\\
    $d_0$ & $5.4h \mbox{ Mpc}$ & Galaxy correlation length
    [Eq.~\eqref{eq:12}]\\
    $\eta$ & $1.77$ & Galaxy correlation slope
    [Eq.~\eqref{eq:12}]\\
    $\alpha$ & $2$ & 
    \begin{minipage}{4truecm}
      \strut Galaxy correlation redshift\\
      \strut dependence [Eq.~\eqref{eq:12}]
    \end{minipage}\\
    $\sigma_W$ & $20''$ & Smoothing length
  \end{tabular}
\end{table}

\begin{figure}[!t]
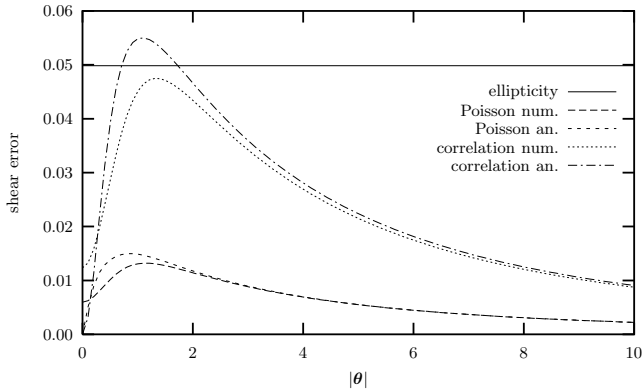

  \begin{center}
    \resizebox{\hsize}{!}{\input fig7.tex}
    \caption{The different contributions to the shear error (i.e.,
      square root of variance) at different radii.  Note that the
      analytical estimates, obtained from the expressions of
      Sect.~\ref{sec:slowly-varying-fields}, are in very good
      agreement with the numerical results.}
    \label{fig:3}
  \end{center}
\end{figure}

\begin{figure}[!t]
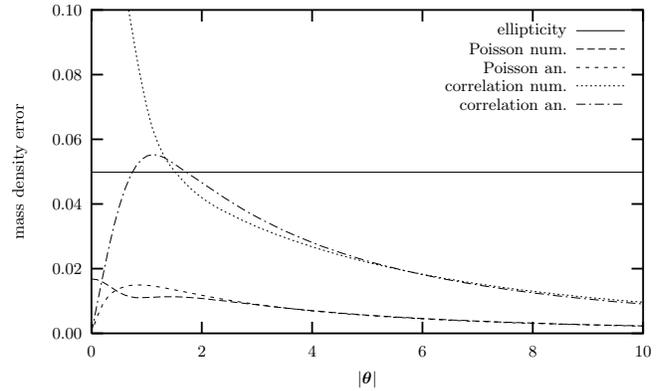

  \begin{center}
    \resizebox{\hsize}{!}{\input fig8.tex}
    \caption{The different contributions to the mass noise at
      different radii.  Although the analytical curves refer to the
      shear noise, they provide a very good estimate for all radii
      larger than $1'$.  The scale of this figure is the same as
      Fig.~\ref{fig:3}; the curve for the numerical evaluation of the
      correlation noise reaches about $0.67$ at the origin.}
    \label{fig:4}
  \end{center}
\end{figure}

Using this lens model, we have carried out the integrations of the
various terms of Eq.~\eqref{eq:26} numerically, and compared the
results obtained with the analytical predictions of
Sect.~\ref{sec:slowly-varying-fields}.  Table~\ref{tab:1} shows the
relevant parameters used in the calculations.  The results obtained
are summarized in Figs.~\ref{fig:3} and \ref{fig:4}.

Figure~\ref{fig:2} shows the lens profile together with the tangential
shear.  In this figure we have also plotted the weight function
$W(\vec\theta)$ used in the calculations.  As shown by
Fig.~\ref{fig:2}, the slowly varying shear approximation is not really
justified in this case, since the lens shear shows significant
deviations from linearity on the scale of the weight function.
However, we find interesting to consider this situation in order to
better test our analytical approximations.

Figure~\ref{fig:3} shows, for different radii, the expected errors on
the shear coming from different sources of noise.  The horizontal line
is the intrinsic ellipticity noise, which clearly is constant on the
whole field.  Note that the analytical estimate for this noise source
is exact.  As shown by Fig.~\ref{fig:3}, correlation of galaxies is
the main source of noise at $\lvert \vec\theta \rvert \simeq 1'$,
while Poisson noise is relatively unimportant (see
Sect.~\ref{sec:impact-observations} below for further comments on this
point).  The analytical approximations given in
Sect.~\ref{sec:slowly-varying-fields} perform surprisingly well, with
deviations below $10\%$ for most radii.  The larger deviations
observed around $\lvert\vec\theta\rvert = 1'$ are expected, since at
this radius the shear deviates significantly from the linear
approximation (the same is true around the origin, where the shear
vanishes).  We also point out that the analytical estimate for the
correlation noise is slightly less accurate than the one for Poisson
noise mainly because Eq.~\eqref{eq:41} is a \textit{zero-order
  estimate\/} (in the sense that terms containing shear derivatives
have not been considered in this equation).

Figure~\ref{fig:4} shows similar results for the noise on the mass map
$\kappa(\vec\theta)$.  The numerical integrations have been performed
using Eq.~\eqref{eq:27}, while the analytical estimates are the same
as for the shear (i.e., we have assumed that the noise on the mass map
is the same as the noise on the shear).  Note that, except at small
radii, the analytical estimates for the shear provide an accurate
measure of the noise on the mass map.  In reality, for more
complicated lenses this property is not verified at the degree of
precision of Fig.~\ref{fig:4}; nevertheless, the analytical
expressions for the shear noise are always a good order-of-magnitude
estimate for the noise in the mass map.

\section{Impact on observations}
\label{sec:impact-observations}

\begin{figure}[!t]
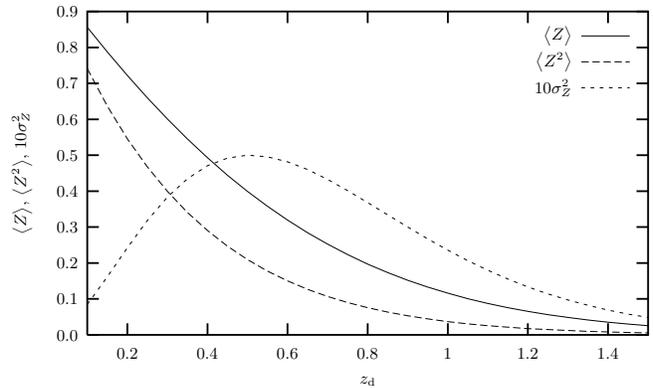

  \begin{center}
    \resizebox{\hsize}{!}{\input fig3.tex}
    \caption{The mean cosmological weight and its moments as a
      function of the lens redshift $z_\mathrm{d}$.  Note that the
      quantity $\sigma_Z$ has a peak around $z_\mathrm{d} = 0.5$.}
    \label{fig:5}
  \end{center}
\end{figure}

So far we have studied the statistical properties of two shear
estimators from a purely analytical point of view .  Clearly, the
interest of our results lies on the consequences of the various noise
sources on weak lensing studies.  In this section we consider the
expressions obtained in Sects.~\ref{sec:slowly-varying-fields} and
\ref{sec:more-accurate-shear} from a more practical point of view,
with particular attention to their dependence on critical
observational parameters (such as the redshift and the strength of the
lens, the density of galaxies, and their mean redshift).

As already stressed, various sources of noise contribute to the
expected error of both estimators.  In particular, we have found three
main contributions:
\begin{enumerate}
\item the intrinsic scatter of source ellipticities;
\item the Poisson noise, both on the positions of galaxies and on
  their redshift;
\item the clustering of galaxies, characterized by the two-point
  correlation function $\xi$.
\end{enumerate}
Depending on some key parameters, these sources of noise can have
different relative importance for the total error of the measured
shear, and thus of the reconstructed lens mass map.

\begin{figure}[!t]
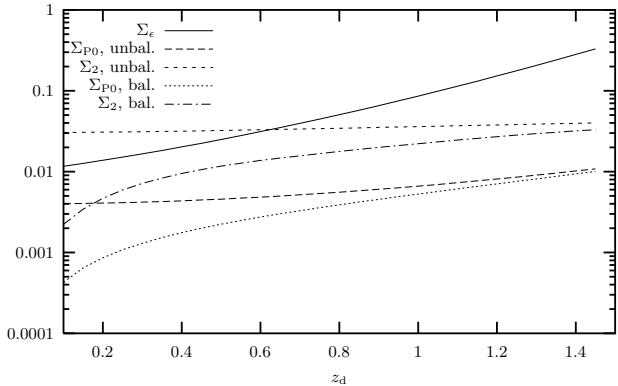

  \begin{center}
    \resizebox{\hsize}{!}{\input fig5.tex}
    \caption{Typical errors on the shear due to various sources of
      noise.  The plot shows the numerical factors in front of some
      equations describing errors on $\hat\gamma$.  For all graphs we
      assumed $\rho = 50 \mbox{ gal arcmin}^{-2}$, $\sigma_W = 1
      \mbox{ arcmin}$, and $\gamma = 0.1$.  For the estimator
      \eqref{eq:4}, Eq.~\eqref{eq:38} gives the error due to the
      intrinsic scatter of source ellipticities $\Sigma_\epsilon$
      (assuming $\sigma_\epsilon = 0.25$), Eq.~\eqref{eq:39} the
      Poisson noise $\Sigma_\mathrm{P0}$ (both in redshift and in
      position), and Eq.~\eqref{eq:42} the noise due to clustering of
      galaxies $\Sigma_2$; for the estimator \eqref{eq:5}, the noise
      due to source ellipticities is still given by Eq.~\eqref{eq:38},
      the Poisson noise (now only on redshift) is given by
      Eq.~\eqref{eq:46}, and the clustering noise by
      Eq.~\eqref{eq:47}.  Note that the relative importance of error
      source depends critically on the shear $\gamma$.  For example,
      for a lens with $\gamma = 0.3$ the clustering noise for the
      balanced estimator would be larger than the intrinsic
      ellipticity noise in the redshift range $0.2 < z_\mathrm{d} <
      0.9$.}
    \label{fig:6}
  \end{center}
\end{figure}

For weak lenses the leading term is the intrinsic scatter of
ellipticities.  In fact, the Poisson noise and the noise due to the
clustering of galaxies are proportional to the shear, and thus are
expected to have a negligible impact on very weak lenses.  We note
that this point is very important for cosmic shear studies.

\begin{table}[b!]
  \centering
  \caption{Noise terms as a function of the observation depth.  The
    table reports the expected noise contributions (square root of
    variance contributions) from various sources of noise.    The
    last column gives the logarithmic slope of the respective noise
    term in terms of $\rho$, i.e.\ the constant $k$ in the relation
    $\Sigma_X \propto \rho^k$.}
  \label{tab:2}
  \begin{tabular}{lcccc}
    Noise term& \multicolumn{3}{c}{Density $\rho$ (arcmin$^{-2}$)} &
    $k$ \\
    & 25 & 50 & 100 &\\
    \hline
    Ellipticity noise $\Sigma_\epsilon$ & 0.025689 & 0.016600 &
    0.010996 & $-0.60$\\
    \multicolumn{5}{c}{\hrulefill\ Unbalanced estimator\ \hrulefill}\\
    Poisson noise $\Sigma_{P0}$ & 0.006039 & 0.004196 & 0.002933 & $-0.52$\\
    Correlation noise $\Sigma_2$ & 0.033567 & 0.031173 & 0.029393 & $-0.10$\\
    \multicolumn{5}{c}{\hrulefill\ Balanced estimator\ \hrulefill}\\
    Poisson noise $\Sigma_{P0}$ & 0.002155 & 0.001300 & 0.000803 & $-0.72$\\
    Correlation noise $\Sigma_2$ & 0.009067 & 0.007152 & 0.005812 & $-0.33$
  \end{tabular}
\end{table}

The noise due to galaxy clustering shows rather surprising properties
if compared to other sources of errors.  As already pointed out in
Sect.~\ref{sec:covariance}, ellipticity and Poisson noise decrease
with the density of background sources as $1/\sqrt{\rho}$; in
contrast, the noise due to clustering is formally independent of the
density of galaxies.  Hence, this source of error is expected to be
particularly important in deep observations and \textit{can represent
  a major limitation for weak lensing studies in the near future}.  In
reality, deeper observations also imply a change of the redshift
distribution $p_z$ and, in particular, an increase of $\langle z
\rangle$.  As a result, the correlation error decreases for deep
observations, but not as fast as other sources of noise.

Table~\ref{tab:2} reports typical figures for the various noise terms
as a function of the density of background galaxies.  Calculations for
that table have been carried out using $\sigma_\epsilon = 0.25$,
$\sigma_W = 1 \mbox{ arcmin}$, and a lens shear $\gamma = 0.1$.  In
order to simulate the increase of the mean redshift for deep
observations, we have used a galaxy redshift probability distribution
$p_z(z)$ of the form of Eq.~\eqref{eq:29}, with $\beta = 1.5$ and
$z_0$ \textit{dependent on the density\/} $\rho$.  In particular, we
have used $z_0 = 0.6$ for $\rho = 25 \mbox{ arcmin}^{-2}$, $z_0 = 0.7$
for $\rho = 50 \mbox{ arcmin}^{-2}$, and $z_0 = 0.8$ for $\rho = 100
\mbox{ arcmin}^{-2}$.  Looking at Table~\ref{tab:2}, we find
surprisingly that the ellipticity noise decreases as $\rho^{-0.60}$
instead of as $\rho^{-1/2}$, as suggested by Eq.~\eqref{eq:38}.  The
$0.1$ increase in the logarithmic slope of $\Sigma_\epsilon$ is due to
the change in the galaxy redshift distribution considered here, and in
particular is related to the coefficient $1/\langle Z \rangle^2$ in
Eq.~\eqref{eq:37}.  In contrast, the Poisson noise $\Sigma_{P0}$ for
the unbalanced estimator almost exactly keeps the $\rho^{-1/2}$
dependence.  This is due to a cancellation effect in
Eq.~\eqref{eq:37}, between $\langle Z \rangle^2$ in the denominator
and $\bigl\langle Z^2 \bigr\rangle$ in the numerator.  The correlation
noise $\Sigma_2$ for the unbalanced estimator is found to be basically
constant with $\rho$; again, the $0.1$ difference between the expected
logarithmic slope ($\Sigma_2 \propto \rho^0$) and the observed one
($\Sigma_2 \propto \rho^{-0.1}$) can be attribute to a $1/\langle Z
\rangle^2$ factor [see Eq.~\eqref{eq:41}].

Turning to the balanced estimator, we immediately note a more
pronounced dependence of the noise terms on $\rho$.  The Poisson noise
decreases as $\rho^{-0.72}$, i.e.\ with a logarithmic slope $0.22$
larger than suggested by Eq.~\eqref{eq:46}.  We can certainly
attribute $0.10$ increase to the factor $1/\langle Z \rangle^2$ in
Eq.~\eqref{eq:45}; the remaining $0.12$ increase is due to
$\sigma_Z^2$, which decreases as the bulk of sources is shifted
towards high redshift.  The correlation noise is found to have a
density dependence $\rho^{-0.33}$.  It is interesting to note that for
both Poisson noise and correlation noise we gain an exponent $-0.2$
with respect to the unbalanced estimator.

For both estimators, the correlation noise is found to be significant
(we recall that calculations for Table~\ref{tab:2} have been carried
out assuming a shear of only $0.1$).  Fortunately, as we will see in
the next section, this source of noise disappears if we have
information on the redshifts of galaxies (e.g., though photometric
redshifts) and if we make use of this data.  A second important point
is the dependence of the noise terms on the lens redshift.
Figure~\ref{fig:6} gives a numerical estimate for the various sources
of error as a function of $z_\mathrm{d}$.  All errors increase with
redshift, and in particular the error due to source ellipticities is
quickly increasing.  In fact, the lensing signal decreases as
$1/\langle Z \rangle$, thus is strongly dependent on the lens redshift
(we recall that $Z(z) = 0$ for $z \leq z_\mathrm{d}$).  Other sources
of noise are less dependent on the redshift of the lens.  It is also
interesting to observe that the largest contribution to Poisson and
clustering noise comes from the angular dependence [compare the curve
$\Sigma_\mathrm{P0}$ for the unbalanced estimator, which is relative
to Poisson noise in angular positions and redshifts of galaxies, with
the curve $\Sigma_\mathrm{P0}$ for the balanced estimator, which
includes only redshift Poisson noise; similarly for the curves
$\Sigma_2$].  This last point is basically due to the ``compression''
of distances in the redshift space.  Assuming that the redshift is
only cosmological, two galaxies at slightly different redshifts (say
$\Delta z = 0.01$) can actually be several Mpc away, so that only weak
correlations are expected for such galaxies.  On the other hand, their
cosmological weight is basically the same, and thus no extra noise on
the shear is expected.  We also stress here the different angular
dependence of the clustering noise, which decreases just as $1/ \lvert
\vec\theta - \vec\theta' \rvert$ at large radii (see
Sect.~\ref{sec:covariance-xi-neq-1}), i.e.\ it introduces a long-range
correlation on the map (in contrast, ellipticity and Poisson noise
correlation decrease very rapidly with $\lvert \vec\theta -
\vec\theta' \rvert$, as a Gaussian).

Finally, it is worth to mention a general technique often used to
estimate errors of weak lensing studies, namely
\textit{bootstrapping}.  This method is based on the generation of
many datasets from the observed data to be used for the determination
of errors.  Suppose that we observe $N$ galaxies with ellipticities
$\bigl\{ \epsilon^{(n)} \bigr\}$ and positions $\bigl\{
\vec\theta^{(n)} \bigr\}$.  We can then construct a new galaxy catalog
by randomly drawing $N$ galaxies from the original catalog,
\textit{with replacement}.  Because of the replacements, we are
allowed to use a galaxy, which is represented here by $(\vec\theta,
\epsilon)$, more that once.  We can then use the new catalog to
perform a weak lensing mass reconstruction.  By repeating the whole
process several times, we eventually obtain the noise properties of
the mass map.  Bootstrapping is a very simple and reasonably robust
method to obtain error estimates, but unfortunately relies on the
assumption of \textit{independent data}.  In our specific case,
because of the spatial correlation of galaxies, we are not really
dealing with independent data, and thus, at least in principle, the
use bootstrapping is not fully justified.  Moreover, the use of
analytical expressions provided here for the noise estimate has the
advantage of showing clearly the different noise contributions and of
allowing us to predict the expected errors without using any specific
dataset.

\section{Sources with known redshifts}
\label{sec:case-source-with}

When some kind of knowledge about the redshift of each galaxy is
available, for example through photometric redshifts, we can take
advantage of the extra information to reduce the error on the shear
(and thus on the mass map) by adopting some improved estimator.

Suppose that \textit{measured\/} redshifts $\bigl\{ \hat z^{(n)}
\bigr\}$ for source galaxies are available, together with their
expected errors.  Then, assuming a particular cosmological model, we
can convert these redshifts into the corresponding cosmological weights
$\bigl\{ \hat Z^{(n)} \bigr\}$.  If the estimates of the redshifts are
\textit{unbiased\/} and their expected errors small, we can take the
cosmological weights unbiased, i.e.\ with some ``symmetric'' scatter
around the true weights.

In order to take into account errors on measured redshifts, we
introduce $p_{\hat z}(\hat z | z)$, the Baysian probability
distribution for $\hat z$.  This function gives the probability of
measuring $\hat z$ when the real redshift is $z$.  Note that we assume
that the probability distribution for $\hat z$ depends only on $z$,
the true redshift.  Referring to photometric redshifts, we observe
that, on one hand, this approximation neglects some key points, such
as the dependence of the error on the photometry (and thus luminosity)
of the galaxy; on the other hand, since, except for the redshift $z$,
we are dealing with quantities which can be taken to be uncorrelated
with the measured redshift $\hat z$ of a galaxy ($\vec \theta$ and
$\epsilon^\mathrm{s}$ have clearly nothing to do with $\hat z$), we
can safely simplify the discussion using $p_{\hat z}(\hat z | z)$.
When performing averages, we need to include the new random variable
$\hat z$.  As a result, we have to modify Eqs.~\eqref{eq:8} and
\eqref{eq:9} with the replacement
\begin{equation}
  \label{eq:52}
  p_z(z) \mapsto \int_0^{\infty} p_{\hat z}(\hat z | z) \, \diff \hat z
  \; .
\end{equation}
Note that in terms of the probability distribution $p_{\hat z}(\hat z
| z)$, the condition of unbiased cosmological weights estimates can be
written as
\begin{equation}
  \label{eq:53}
  \langle \hat Z \rangle \equiv \int p_{\hat z}(\hat z | z) Z(\hat z)
  \, \diff \hat z = Z(z) \; .
\end{equation}
In order simplify the discussion, in the following we will assume that
the expected error on the cosmological weight is a simple function of
the redshift of the galaxy, which we write $\sigma_{\hat Z}(z)$:
\begin{equation}
  \label{eq:54}
  \sigma_{\hat Z}(z) = \int_0^\infty p_{\hat z}(\hat z | z) \bigl[
  Z(\hat z) - Z(z) \bigr]^2 \, \diff \hat z \; .
\end{equation}
We will often refer to the variance of measured redshifts.  This
quantity is defined as
\begin{equation}
  \label{eq:55}
  \sigma^2_{\hat Z} \equiv \int_0^\infty p_z(z) \sigma_{\hat Z}(z)
  \, \diff z \; .
\end{equation}
We stress that this variance depends on both the errors in redshifts and
on the source redshift probability distribution.  To give an example
of the precision that can be reached by photometric redshifts when
accurate photometry is available, we note that for the Hubble Deep
Field North a scatter of the order of $0.07 (1 + z)$, without outliers
and bias, has been reached \citep{2000ApJ...536..571B}.

We will now investigate two possible uses of the quantities $\bigl\{
\hat Z^{(n)} \bigr\}$ for shear estimators.  Our starting model will
be the estimator of Eq.~\eqref{eq:5}, since the estimator
\eqref{eq:4} has already been proven to have a larger variance.

\subsection{Unweighted estimator}
\label{sec:unweighted-estimator}

In Sect.~\ref{sec:more-accurate-shear} we have seen that the use of a
balanced estimator on the source positions completely eliminates
Poisson noise due to the galaxy positions and significantly reduces
the correlation noise.  Since estimates for the cosmological weights
are now available, we can now think of replacing the average $\langle
Z \rangle$ in the denominator of Eq.~\eqref{eq:5} with the measured
values for the cosmological weight.  This leads to the new shear
estimator
\begin{equation}
  \label{eq:56}
  \hat\gamma(\vec\theta) = \dfrac{\sum_{n=1}^N \epsilon^{(n)}
    W \bigl( \vec\theta - \vec\theta^{(n)} \bigr)}{\sum_{n=1}^N \hat
    Z^{(n)} W \bigl( \vec\theta - \vec\theta^{(n)} \bigr)} \; .  
\end{equation}
This estimator is called ``unweighed'' since we are not taking
advantage of the redshift information to weight galaxies (a weighted
estimator will be discussed below in
Sect.~\ref{sec:weighted-estimator}). Using a technique very similar to
the one discussed above, we can study the statistical properties of
this estimator, and in particular its mean and variance.

Regarding the mean, in the linear approximation this estimator behaves
exactly as the estimators considered so far, i.e.\ Eq.~\eqref{eq:15}
holds.  In the slowly varying shear approximation, moreover, we have
$\bigl\langle \hat\gamma(\vec\theta) \bigr\rangle =
\gamma(\vec\theta)$.  We stress that a key point in deriving these
results is Eq.~\eqref{eq:53}, i.e.\ the condition of unbiased
cosmological weights.

Using the slowly varying shear approximations and in case of a
vanishing two-point galaxy correlation function, the covariance of
$\hat\gamma$ is
\begin{align}
  \label{eq:57}
  &\Cov_{ij}(\hat\gamma; \vec\theta, \vec\theta') = \frac{1}{4 \pi
    \langle Z \rangle^2 \rho \sigma_W^2} \biggl[ \sigma^2_\epsilon
  \delta_{ij} + \sigma^2_{\hat Z} \gamma_i(\vec\theta)
  \gamma_j(\vec\theta') \notag\\
  & \qquad {} + \biggl( \frac{\bigl\langle Z^2 \bigr\rangle
    \sigma_W^2}{2} \delta_{i'j'} - \frac{\langle Z \rangle^2}{4}
  \varphi_{i'} \varphi_{j'} \biggr) \gamma_{i,i'} (\bar{\vec\theta})
  \gamma_{j,j'}
  (\bar{\vec\theta})  \biggr] \notag\\
  & \quad {} \times \exp \left( - \frac{\lvert \vec\theta -
      \vec\theta' \rvert^2}{4 \sigma_W^2} \right) \; ,
\end{align}
where $\sigma^2_{\hat Z}$ is defined in Eq.~\eqref{eq:55}.  Note that
Eq.~\eqref{eq:57} is very similar to Eq.~\eqref{eq:45}, with the
replacement $\sigma^2_Z \mapsto \sigma^2_{\hat Z}$.  In general
$\sigma^2_{\hat Z}$ is smaller than $\sigma^2_Z$, since a redshift
estimate will hopefully provide a better constraint on the galaxy
cosmological weights than using a straight average for all galaxies.
We stress, moreover, that $\sigma^2_{\hat Z}$ is mainly set by the
redshift inaccuracies, since it represents the scatter of a
\textit{measured\/} cosmological weight with respect to the true
value, while $\sigma^2_Z$ depends only on the galaxy redshift
distribution.  Clearly, if no errors on the redshift estimation are
present we have $\sigma^2_{\hat Z} = 0$.

Calculations for the case of a non-vanishing galaxy two-point
correlation function, not reported here, are rather surprising.  In
fact, using the approximations adopted in
Sect.~\ref{sec:covariance-xi-neq}, we obtain a vanishing contribution
to the covariance.  In other words, \textit{the availability of some
  redshift information completely cancels the noise due to the
  redshift correlation of background galaxies.}  In deriving this
result, a key role is played by the assumption that the measured
cosmological weights are unbiased.

\subsection{Weighted estimator}
\label{sec:weighted-estimator}

The use of an unweighted estimator is rather convenient for the
discussion of the noise properties of the shear, but is not optimal.
In fact, we can take further advantage of the knowledge of the
redshifts of galaxies by suppressing galaxies which are (or are taken
to be) in front of the cluster and enhancing distant galaxies, which
are well affected by the lens.  A simple investigation shows that the
weight to use for each galaxy is proportional to $Z^{(n)}$.  As a
result, we can think to use the shear estimator \citep[see
also][]{1999A&A...342..337L}
\begin{equation}
  \label{eq:58}
  \hat\gamma(\vec\theta) = \dfrac{\sum_{n=1}^N \epsilon^{(n)}
    \hat Z^{(n)} W \bigl( \vec\theta - \vec\theta^{(n)}
    \bigr)}{\sum_{n=1}^N \bigl( \hat Z^{(n)} \bigr)^2 W \bigl(
    \vec\theta - \vec\theta^{(n)} \bigr)} \; .
\end{equation}
Unfortunately, this estimator turns out to be \textit{biased\/} if the
estimated cosmological weights $\hat Z^{(n)}$ are not identical to the
true ones $Z^{(n)}$.  In fact we have
\begin{equation}
  \label{eq:59}
  \left\langle \hat\gamma(\vec\theta) \right\rangle =
  \frac{\bigl\langle Z^2 \bigr\rangle}{\bigl\langle \hat Z^2
  \bigr\rangle} \int_\Omega W(\vec\theta - \vec\theta')
  \gamma(\vec\theta') \, \diff^2 \theta \; .
\end{equation}
In other words, we will underestimate the shear by a factor $1 +
\sigma_{\hat Z}^2 / \langle Z \rangle^2$.  This, hopefully, is a
number very close to one, so we will for the moment ignore this effect
(for example, assuming $\sigma_{\hat Z}^2 < \sigma_Z^2$, this factor
differs from $1$ by an amount smaller than $0.02$ for a lens at
redshift $z_\mathrm{d} = 0.3$).

The expression for the covariance of $\hat \gamma$ is in general
rather complicated because of the presence of complex combinations of
averages of $Z$ and $\hat Z$.  For this reason, we do not explicitly
write here the general expression.  In the special case of no errors
(so that $\hat Z = Z$), however, we obtain the simple result
\begin{align}
  \label{eq:60}
  & \Cov_{ij}(\hat\gamma; \vec\theta, \vec\theta') = \frac{1}{4 \pi
    \bigl\langle Z^2 \bigr\rangle \rho \sigma_W^2} \biggl[
  \sigma^2_\epsilon \delta_{ij} \notag\\
  & \qquad {} + \biggl( \frac{\bigl\langle Z^2 \bigr\rangle
    \sigma_W^2}{2} \delta_{i'j'} - \frac{\langle Z \rangle^2}{4}
  \varphi_{i'} \varphi_{j'} \biggr) \gamma_{i,i'} (\bar{\vec\theta})
  \gamma_{j,j'} (\bar{\vec\theta})  \biggr] \notag\\
  & \quad {} \times \exp \left( - \frac{\lvert \vec\theta -
      \vec\theta' \rvert^2}{4 \sigma_W^2} \right) \; ,
\end{align}
This equation should be compared with Eq.~\eqref{eq:57} with
$\sigma_{\hat Z}^2 = 0$.  We see that the main difference is the
factor $\bigl\langle Z^2 \bigr\rangle$ which replaces $\langle Z
\rangle^2$.  We can thus conclude that, in case of complete redshift
knowledge, the estimator \eqref{eq:58} performs better than the one of
Eq.~\eqref{eq:56}.  Actually, the gain is rather low (about $10\%$ for
the lens at redshift $z_\mathrm{d} = 0.3$), and one could also prefer
to keep the unweighted estimator \eqref{eq:56} rather than the
weighted one as long as the lens redshift is much smaller than the
median redshift of galaxies.

\subsection{Discussion}
\label{sec:discussion}

As shown above, one can well take advantage of redshift information to
reduce the noise of shear estimates.  Since weak lensing analysis
normally deals with thousands of galaxies, redshift information is
generally available only though photometric redshifts, and thus errors
on redshifts must be expected.

In general, galaxy redshifts, even with errors, are very valuable data:
They can be used to severely reduce the noise of the mass
reconstruction, especially the contribution due to galaxy
correlations; moreover, they can also be used to select background
galaxies or, more generally, to weight galaxies depending on the
lensing signal that they carry.  On the other hand, some care should
be used when photometric redshifts are used to select galaxies.  In
fact, errors on photometric redshifts are extremely sensitive to the
accuracy of the photometry and on the bands used.  Large errors,
clearly, imply some kind of ``contamination'' between foreground and
background galaxies and, ultimately, introduce a bias on the shear
estimate [see above Eq.~\eqref{eq:59}].

In general, we find that photometric redshifts can almost always be
used together with the estimator \eqref{eq:56} (which, we recall, can
be significantly less noisy than estimators not accounting for galaxy
redshifts).  We think, instead, that the estimator \eqref{eq:58},
should be used only if errors on the redshifts are statistically well
known and not too large or, alternatively, if the lens is at high
redshift.  In fact, for a high redshift lens, we expect a significant
fraction of foreground galaxies: The use of a ``filter,'' provided by
the weight $Z^\mathrm{(n)}$ in the numerator of Eq.~\eqref{eq:58}, can
significantly reduce the noise of the shear and thus of the mass map.

\section{Conclusions}
\label{sec:conclusions}

In this paper we have presented a careful statistical analysis of the
shear estimators most widely used in weak lensing mass
reconstructions.  In particular, we have obtained analytical
expressions for the bias and the covariance of the estimators, and
identified the various contributions to the noise, as summarized in
the following items.
\begin{itemize}
\item Intrinsic spread of ellipticities.  This term, which is common
  to all estimators, is basically the only one considered by previous
  investigations \citep[see][]{1995ApJ...449..460K,
    1998A&A...335....1L, 2000MNRAS.313..524W}.
\item Poisson noise.  Poisson noise is due to the spread of galaxies
  in positions and redshift.  We have shown that a simple, balanced
  estimator is able to eliminate the angular component of Poisson
  noise, thus reducing this source of error by a factor $\sim 4$.  If
  redshift information is available, redshift Poisson noise can also
  be reduced using suitable estimators.
\item Correlation noise.  The clustering of galaxies introduces a
  further source of noise.  We have shown that this is generally the
  leading noise contribution, even for the balanced estimator.  This
  source of noise, in contrast to other sources, does not formally
  decrease with the density of background galaxies.  This error can be
  severely depressed if redshift information is available.
\end{itemize}
We have also discussed the dependence of these noise terms on several
relevant parameters, such as the lens redshift, the lens strength, the
density of background galaxies, and the smoothing length used.
Finally, we have tested our analytical framework by performing
numerical integrations in a specific model, and by comparing the
results obtained with our analytical predictions.  We have then proved
that our finite-form expressions for the shear noise are correct to
within $10\%$; moreover, we have shown that these expressions can be
safely taken also as good approximations for the error on the mass
map.

\acknowledgements We would like to thank Giuseppe Bertin for helpful
discussions.

\appendix
  
\section{Linear approximation}
\label{sec:linear-approximation-1}

For the study of the properties of the estimator \eqref{eq:5} we have
used a standard technique often applied in statistics \citep[][see,
e.g.]{Eadie}. Suppose that $X$ is a multidimensional random variable
and $Y = f(X)$, with $f$ a smooth function.  If $X$ has small
covariance, we can expand $f$ to first-order and obtain this way a
simple expression for the mean and the covariance of $Y$:
\begin{gather}
  \label{eq:61}
  \langle Y \rangle \simeq f \bigl( \langle X \rangle \bigr) \; , \\
  \label{eq:62}
  \Cov_{ij}(Y) \simeq C_{ii'} C_{jj'} \Cov_{i'j'}(X) \; , \\
  \intertext{where}
  \label{eq:63}
  C_{ij} = \left. \frac{\partial f_i}{\partial X_j} \right\rvert_{X =
    \langle X \rangle} \; .
\end{gather}
The goodness of the first-order approximation can be verified by using
a second-order expansion.  Calculations are particularly simple for
the average value:
\begin{equation}
  \label{eq:64}
  \langle Y \rangle \simeq f \bigl( \langle X \rangle \bigr) +
  \frac{1}{2} \frac{\partial^2 f}{\partial X_1 \partial X_j}
  \Cov_{ij}(X) \; .
\end{equation}
The second term represents the second-order correction and can be used
as an estimate for the error made.

\section{Detailed statistical analysis}
\label{sec:deta-stat-analys}

In this appendix we will carry out in some detail a statistical
analysis for the shear estimator \eqref{eq:5}, thus recovering the
results stated in Sect.~\ref{sec:more-accurate-shear}.

\subsection{Average value}
\label{sec:average-value-1}

The average value of the estimator is easily evaluated using the
linear approximation.  In particular, we have
\begin{gather}
  \label{eq:65}
  \langle N \rangle = \rho \langle Z \rangle \int_\Omega
  W(\vec\theta - \vec\theta') \gamma(\vec\theta') \, \diff^2 \theta'
  \; , \\
  \label{eq:66}
  \langle D \rangle = \rho \langle Z \rangle \; .
\end{gather}
As a result, using Eq.~\eqref{eq:61}, we find
\begin{equation}
  \label{eq:67}
  \langle \hat\gamma(\vec\theta) \rangle = \int_\Omega
  W(\vec\theta - \vec\theta') \gamma(\vec\theta') \, \diff^2
  \theta' \; .
\end{equation}

\subsection{Covariance}
\label{sec:covariance-1}

Calculations for the covariances are more difficult.  The scatter of
the numerator from the average value can be written as
\begin{align}
  \label{eq:68}
  N_i - \langle N_i \rangle = \sum_{n=1}^N \biggl[ & \Bigl(
  \epsilon_i^{\mathrm{s}(n)} + \gamma_i \bigl( \vec\theta^{(n)} \bigr)
  Z^{(n)} \Bigr) W\bigl( \vec\theta - \vec\theta^{(n)} \bigr) \notag\\
  & {} - \frac{\langle Z \rangle}{A} \int_\Omega W(\vec\theta -
  \vec\phi) \gamma_i(\vec\phi) \, \diff^2 \phi \biggr] \;
  .
\end{align}
For the denominator we have
\begin{equation}
  \label{eq:69}
  D - \langle D \rangle = \langle Z \rangle \sum_{n=1}^N \biggl[ W\bigl(
  \vec\theta - \vec\theta^{(n)} \bigr) - \frac{1}{A} \biggr] \; .
\end{equation}
Using Eq.~\eqref{eq:68}, we can write the covariance $\Cov(N_i N'_j)$
as
\begin{alignat}{2}
  \label{eq:70}
  \Cov&\hbox to 0pt{$(N_i N'_j; \vec\theta, \vec\theta') = {} $}\notag\\
  \Bigg\langle & \sum_{n=1}^N \biggl[ && \!
  \Bigl( \epsilon_i^{\mathrm{s}(n)} + \gamma_i \bigl( \vec\theta^{(n)}
  \bigr) Z^{(n)} \Bigr) W\bigl( \vec\theta - \vec\theta^{(n)}
  \bigr) \notag\\
  &&& {} - \frac{\langle Z \rangle}{A} \int_\Omega W(\vec\theta -
  \vec\phi) \gamma_i(\vec\phi) \, \diff^2 \phi \biggr]
  \times {} \notag\\
  & \! \sum_{m=1}^N \biggl[ &&
  \Bigl( \epsilon_j^{\mathrm{s}(m)} + \gamma_j \bigl( \vec\theta^{(m)}
  \bigr) Z^{(m)} \Bigr) W\bigl( \vec\theta' - \vec\theta^{(m)}
  \bigr) \notag\\
  &&& {} - \frac{\langle Z \rangle}{A} \int_\Omega W(\vec\theta' -
  \vec\phi') \gamma_j(\vec\phi') \, \diff^2 \phi' \biggr]
  \Biggr\rangle\; .
\end{alignat}
Similarly to Eq.~\eqref{eq:21}, it is convenient here to write
$\Cov(N_i N'_j) = K_1 + K_2$, where $K_1$ includes terms for which $n
= m$, and $K_2$ terms for which $n \neq m$.

If $n = m$, Eq.~\eqref{eq:8} can be used, leading to
\begin{align}
  \label{eq:71}
  K_1 = {} & \rho \sigma^2_\epsilon \delta_{ij} \int_\Omega
  W(\vec\theta - \vec\phi) W(\vec\theta' - \vec\phi) \,
  \diff^2 \phi
  \notag\\
  & {} + \bigl\langle Z^2 \bigr\rangle \rho \int_\Omega
  \gamma_i(\vec\phi) \gamma_j(\vec\phi) W(\vec\theta -
  \vec\phi) W(\vec\theta' - \vec\phi) \, \diff^2 \phi \; ,
\end{align}
where terms which vanish in the limit $A \rightarrow \infty$ have
been discarted.  If $n \neq m$ we have
\begin{align}
  \label{eq:72}
  K_2 = {} & \rho^2 \int_\Omega \! \diff^2 \phi W(\vec\theta -
  \vec\phi) \gamma_i(\vec\phi) \int_\Omega \!  \diff^2 \phi'
  W(\vec\theta' - \vec\phi') \gamma_j(\vec\phi')
  \notag\\
  & \int_0^\infty \!\!\!\! p_z(z) Z(z) \, \diff z \int_0^\infty
  \!\!\!\! p_z(z') Z(z') \xi(\vec\phi - \vec\phi', z, z') \, \diff z'
  \; .
\end{align}
We recall [see Eq.~\eqref{eq:25}] that this integral involving the
two-point correlation function has been denoted as $\Xi\bigl[ W
\gamma_i Z W' \gamma_j' Z' \bigr]$.  Finally we find for
$\Cov(N_i N'_j)$
\begin{align}
  \label{eq:73}
  & \Cov(N_i N'_j; \vec\theta, \vec\theta') = \rho \sigma^2_\epsilon
  \delta_{ij} \int_\Omega W(\vec\theta - \vec\phi) W(\vec\theta' -
  \vec\phi) \, \diff^2 \phi \notag\\
  & \qquad {} + \bigl\langle Z^2 \bigr\rangle \rho \int_\Omega
  \gamma_i(\vec\phi) \gamma_j(\vec\phi) W(\vec\theta - \vec\phi)
  W(\vec\theta' - \vec\phi) \, \diff^2 \phi \notag\\
  & \qquad {} + \rho^2 \Xi\bigl[ W \gamma_i Z W' \gamma_j' Z' \bigr]
  \; .
\end{align}

Covariances for the other terms can be calculated in a similar manner.
The final result obtained for $\Cov(N_i D')$ is
\begin{align}
  \label{eq:74}
  \Cov (N_i D'; \vec\theta, \vec\theta') = {} & \langle Z \rangle^2
  \rho \int_\Omega W(\vec\theta - \vec\phi) W(\vec\theta' - \vec\phi)
  \gamma_i(\vec\phi) \, \diff^2 \phi \notag\\
  &{} + \langle Z \rangle \rho^2 \Xi\bigl[ W \gamma_i Z W' \bigr] \; .
\end{align}
The result for $\Cov(D N'_j)$ is analogous and will not be written
explicitely.  Finally, the last covariance needed is
\begin{align}
  \label{eq:75}
  \Cov (D D'; \vec\theta, \vec\theta') = {} & \langle Z \rangle^2 \rho
  \int_\Omega W(\vec\theta - \vec\phi) W(\vec\theta' -
  \vec\phi) \, \diff^2 \phi \notag\\
  &{} + \langle Z \rangle^2 \rho^2 \Xi\bigl[ W W' \bigr] \; .
\end{align}

In principle, we could now evaluate the covariance of $\hat\gamma$
using the results obtained so far.  In practice, the resulting
expression is by far too complex to be of any practical use at this
stage.  We thus avoid to write this expression in the general case,
but rather we will consider the slowly varying shear approximation in
the next appendix.

\section{Slowly varying shear}
\label{sec:slowly-vary-fields-1}

We will derive here the slowly varying shear expressions for the
results obtained in Appendix~\ref{sec:deta-stat-analys}.  Again, only
the estimator \eqref{eq:5} will be considered; calculations for the
estimator \eqref{eq:4} are very similar.

\subsection{Average value}
\label{sec:average-value-2}

To first order, the average value of $\hat\gamma$ is just the ratio
between $\langle N \rangle$ and $\langle D \rangle$ [see
Eq.~\eqref{eq:61}].  Let us focus on $\langle N \rangle$, which is
given by Eq.~\eqref{eq:65}.  Inserting Eq.~\eqref{eq:32} in
Eq.~\eqref{eq:65} and using the symmetry of $W$ we get
\begin{equation}
  \label{eq:76}
  \langle N \rangle = \rho \langle Z \rangle \gamma(\vec\theta)
  \int_\Omega W(\vec\theta - \vec\theta') \, \diff^2 \vec\theta' =
  \rho \langle Z \rangle \gamma(\vec\theta) \; .
\end{equation}
Finally, using Eq.~\eqref{eq:61} we obtain
\begin{equation}
  \label{eq:77}
  \langle \hat\gamma(\vec\theta) \rangle = \gamma(\vec\theta) \; .
\end{equation}

\subsection{Covariance ($\xi = 0$)}
\label{sec:covariance-xi-=}

Calculations for the covariance can be similarly carried out using an
expansion for $\gamma$ around the mid-point $\vecbar\theta =
(\vec\theta + \vec\theta') / 2$ (see Fig.~\ref{fig:1}).  This is
convenient because with this choice several linear terms containing
partial derivatives of $\gamma$ vanish.  Hence we have
\begin{align}
  \label{eq:78}
  & \Cov_{ij}(\hat\gamma; \vec\theta, \vec\theta') = \frac{1}{\rho}
  \biggl[ \frac{\sigma^2_\epsilon}{\langle Z \rangle^2} \delta_{ij} +
  \frac{\bigl\langle Z^2 \bigr\rangle}{\langle Z \rangle^2}
  \gamma_i(\vecbar\theta) \gamma_j(\vecbar\theta) \notag\\
  & \quad \qquad {} - \gamma_i(\vec\theta) \gamma_j(\vecbar\theta) -
  \gamma_i(\vecbar\theta) \gamma_j(\vec\theta') + \gamma_i(\vec\theta)
  \gamma_j(\vec\theta') \biggr] \notag\\
  & \qquad {} \times (W \star W)(\vec\theta - \vec\theta') \notag\\
  & \quad {} + \frac{\bigl\langle Z^2 \bigr\rangle}{\langle Z
    \rangle^2 \rho} \gamma_{i,i'} (\bar{\vec\theta}) \gamma_{j,j'}
  (\bar{\vec\theta}) \notag\\
  & \qquad {} \times \int_\Omega W(\vec\theta - \vec\phi)
  W(\vec\theta' - \vec\phi) (\phi_{i'} - \bar\theta_{i'}) (\phi_{j'} -
  \bar\theta_{j'}) \, \diff^2 \phi \; .
\end{align}
We can apply once more the slowly varying shear approximation to this
result.  In the case where $W$ is a Gaussian the final result can be
written in a simpler form:
\begin{align}
  \label{eq:79}
  & \Cov_{ij}(\hat\gamma; \vec\theta, \vec\theta') = \frac{1}{4 \pi
    \langle Z \rangle^2 \rho \sigma_W^2} \biggl[ \sigma^2_\epsilon
  \delta_{ij} + \sigma^2_Z \gamma_i(\vecbar\theta)
  \gamma_j(\vecbar\theta) \notag\\
  & \qquad {} + \biggl( 
  \frac{\bigl\langle Z^2 \bigr\rangle \sigma_W^2}{2} \delta_{i'j'}
  - \frac{\langle Z \rangle^2}{4} \varphi_{i'}
    \varphi_{j'} \biggr) \gamma_{i,i'} (\bar{\vec\theta}) \gamma_{j,j'}
  (\bar{\vec\theta})  \biggr] \notag\\
  & \quad {} \times \exp \left( - \frac{\lvert \vec\theta
      - \vec\theta' \rvert^2}{4 \sigma_W^2} \right) \; ,
\end{align}
i.e.\ Eq.~\eqref{eq:45}.

\subsection{Check of the linear approximation}
\label{sec:check-line-appr}

We can explicitely verify the goodness of the first-order
approximation by checking the second-order terms of Eq.~\eqref{eq:64}.
These terms are both of the form
\begin{equation}
  \label{eq:80}
  \frac{\gamma(\vec\theta)}{\rho} \biggl[ \int_\Omega W(\vec\theta')
  W(\vec\theta') \, \diff^2 \theta' \biggr] =
  \frac{\gamma(\vec\theta)}{\rho A_W} \; .
\end{equation}
The two terms have opposite sign.  In other words, the average value
of $\hat\gamma$ obtained above is correct to second order [actually it
is easy to show that the estimator \eqref{eq:5} really is unbiased to
all orders].  We also note that the \textit{relative\/} contribution
of this terms is of the order $1/(\rho A_W) \sim 1/N_W$, where, we
recall, $N_W$ the expected number of galaxies inside the effective
area of $W$.  Note that $1/N_W$ is also the expected relative variance
of the denominator $D$ of Eq.~\eqref{eq:5}.  We have thus recovered
the condition needed to perform the linear approximation, already
discussed at the end of Sect.~\ref{sec:linear-approximation}.

\subsection{Covariance ($\xi \neq 0$)}
\label{sec:covariance-xi-neq-1}

Given the difficulties encountered to study the contribution to the
covariance from the two-point correlation function, we will carry out
the calculations only in the case where $W$ is a Gaussian of the form
\eqref{eq:3}.

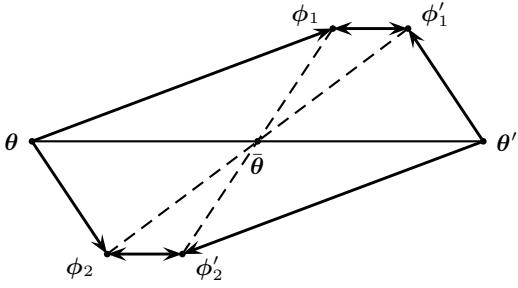
\begin{figure}[t!]
  \begin{center}
    \psset{unit=1cm}
    \begin{pspicture}(-3.5,-2)(3.5,2)
      \psdots(-3,0)(3,0)(-2,-1.5)(-1,-1.5)(2,1.5)(1,1.5)(0,0)
      \psset{linewidth=0.8pt}
      \psline(-3,0)(3,0)
      \psline[linestyle=dashed](-2,-1.5)(2,1.5)
      \psline[linestyle=dashed](-1,-1.5)(1,1.5)
      \psset{linewidth=1.1pt}
      \psline{->}(-3,0)(1,1.5)
      \psline{->}(3,0)(2,1.5)
      \psline{->}(-3,0)(-2,-1.5)
      \psline{->}(3,0)(-1,-1.5)
      \psline{<->}(-1,-1.5)(-2,-1.5)
      \psline{<->}(1,1.5)(2,1.5)
      \uput[180](-3,0){$\vec\theta$}
      \uput[0](3,0){$\vec\theta'$}
      \uput[30](2,1.5){$\vec\phi_1'$}
      \uput[150](1,1.5){$\vec\phi_1$}
      \uput[210](-2,-1.5){$\vec\phi_2$}
      \uput[-30](-1,-1.5){$\vec\phi_2'$}
      \uput[-90](0,0){$\bar{\vec\theta}$}
    \end{pspicture}
    \caption{Similarly to Fig.~\ref{fig:1}, this graph shows that
      linear terms of the expansion of $\gamma$ around
      $\bar{\vec\theta}$ do not contribute to the term $C_2$.  In
      fact, the integrand evaluated at the upper configuration
      $(\vec\phi_1, \vec\phi_1')$ is the opposite of the integrand
      evaluated at lower configuration $(\vec\phi_2, \vec\phi_2')$.}
    \label{fig:7}
  \end{center}
\end{figure}

Let us focus on a specific term involving the two-point correlation
function, for example the last term of Eq.~\eqref{eq:73}, i.e.\ 
$\Xi_{ij} \equiv \Xi\bigl[ W \gamma_i Z W' \gamma_j' Z' \bigr]$ [in
this section, for simplicity, we will use the symbols $\Xi_{ij}$,
$\Xi_i$, and $\Xi$ to denote, respectively, the last terms of
Eqs.~\eqref{eq:73}, \eqref{eq:74}, and \eqref{eq:75}].  We will use
the expression \eqref{eq:12} for the function $\xi$ with the
approximate exponent $\eta = 2$.  We then write
\begin{equation}
  \label{eq:81}
  \xi(\vec\delta, z, z') = \frac{d_0^2 (1 + z)^\alpha}{D_A^2(z) \lvert
  \vec\delta \rvert^2 + \bigl( D_P(z) - D_P(z') \bigr)^2} \; ,
\end{equation}
where $\vec\delta = \vec\phi - \vec\phi'$ must be expressed in
radians.  In this equation $D_A(z)$ and $D_P(z)$ are, respectively,
the angular diameter distance and the proper distance of an object at
redshift $z$.  The length-scale of $\xi$ is set by $d_0$, and
typically is a few Mpc.  As a result, the two-point correlation
function vanishes unless $z'$ is \textit{very\/} close to $z$ (this
property will be repeatedly used below).  On the other hand, all other
functions involving redshifts, namely $D_A$, $D_P$ (but also $p_z$ and
$Z$), are slowly varying with $z$.  We thus can approximate $\xi$ as
\begin{equation}
  \label{eq:82}
  \xi(\vec\delta, z, z') \simeq \frac{d_0^2 (1 + z)^\alpha}{D_A^2(z)
  \lvert \vec\delta \rvert^2 + \bigl[ D_P'(z) \bigr]^2 (z' - z)^2} \;
  . 
\end{equation}
Moreover, since $\xi$ is basically zero for $z' \neq z$, we can take
as constant the term $p_z(z') Z(z')$ in Eq.~\eqref{eq:23}, and perform
directly the integration over $z'$.  We get then
\begin{align}
  \label{eq:83}
  \Xi_{ij} \simeq {}\! & \int_\Omega \! \diff^2 \phi W(\vec\theta -
  \vec\phi) \gamma_i(\vec\phi) \int_\Omega \!  \diff^2
  \phi' W(\vec\theta' - \vec\phi') \gamma_j(\vec\phi') \notag\\
  & \int_0^\infty \!\!\!\! \bigl[ p_z(z) Z(z) \bigr]^2 \, \diff z
  \int_0^\infty \!\!\!\! \xi(\vec\phi - \vec\phi', z, z') \, \diff z'
\end{align}
Using Eq.~\eqref{eq:82} the last integration is trivial:
\begin{equation}
  \label{eq:84}
  \int_0^\infty \xi(\vec\delta, z, z') \, \diff z' =
  \frac{d_0^2 (1 + z)^\alpha}{D_A(z) D'_P(z) \lvert \vec\delta \rvert}
  \Bigl[ \arctan t \Bigr]^\infty_{t_0} \; ,
\end{equation}
with
\begin{equation}
  \label{eq:85}
  t_0 = - \frac{D'_P(z) z}{D_A(z) \lvert \vec\delta \rvert} \; .
\end{equation}
In normal conditions $t_0$ has always a very large negative value
(unless $z$ is exceedingly small; these cases, however, are of no
interest in weak lensing studies).  We thus can safely replace
$t_0$ with $-\infty$ in Eq.~\eqref{eq:84}, thus obtaining
\begin{equation}
  \label{eq:86}
  \int_0^\infty \xi(\vec\delta, z, z') \, \diff z' \simeq
  \frac{\pi d_0^2 (1 + z)^\alpha}{D_A(z) D'_P(z) \lvert \vec\delta
  \rvert} \equiv \zeta(\vec\delta, z) \; .
\end{equation}
We now can perform the two integrals on $\vec\theta$ and
$\vec\theta'$.  Since the results obtained so far depends only on
$\vec\delta = \vec\phi' - \vec\phi$, we have to evaluate two
``quasi-convolutions,'' the ``quasi'' being due to the fact that the
integrand also contain terms like $\gamma_i(\vec\phi)$ and
$\gamma_j(\vec\phi')$.  Hence we can apply the slowly varying field
approximation (see Fig.~\ref{fig:7}), obtaining the integral
\begin{align}
  \label{eq:87}
  \Xi_{ij} \simeq {} & \gamma_i(\vecbar\theta)
  \gamma_j(\vecbar\theta) \int_0^\infty \bigl[ p_z(z) \bigr]^2
  \bigl[ Z(z) \bigr]^2 \, \diff z \int_\Omega W(\vec\theta -
  \vec\phi) \, \diff^2 \phi \notag\\
  &\int_\Omega W(\vec\theta' - \vec\phi') \zeta(\vec\phi' - \vec\phi,
  z) \, \diff^2 \phi \; .
\end{align}
Using the properties of convolutions we can recast the two integrals
on the angular variables in the form
\begin{equation}
  \label{eq:88}
  \nu(\vec\varphi, z) = \int_\Omega \zeta(\vec\phi, z) (W \star
  W)(\vec\varphi - \vec\phi) \, \diff^2 \vec\phi \; ,
\end{equation}
where $\vec\varphi = \vec\theta' - \vec\theta$.  Using the Gaussian
weight function \eqref{eq:3} and the result obtained above for
$\zeta$, we can perform the integration in polar coordinates.  The
result is
\begin{equation}
  \label{eq:89}
  \nu(\vec\varphi, z) = \frac{\pi^{3/2} d_0^2 (1 + z)^\alpha}{2 D_A(z)
    D_P'(z) \sigma_W} I_0 \left( \frac{\lvert \vec\varphi \rvert^2}{8
    \sigma_W^2} \right) \exp \left( - \frac{\lvert \vec\varphi
    \rvert^2}{8 \sigma_W^2} \right) \; ,
\end{equation}
where $I_0$ is the modified Bessel function of first kind.  The final
result for the covariance term is thus
\begin{align}
  \label{eq:90}
  \Xi_{ij} = {}& \frac{\pi^{3/2} d_0^2}{2 \sigma_W}
  \gamma_i(\vecbar\theta) \gamma_j(\vecbar\theta) \, \eta \left(
    \frac{\lvert \vec\varphi \rvert^2}{8 \sigma_W^2} \right) \notag\\
  & {} \times \int_{z_d}^\infty \frac{(1 + z)^\alpha \bigl[p_z(z) Z(z)
    \bigr]^2}{D_A(z) D_P'(z)} \, \diff z \; .
\end{align}
The function $\eta(x) = I_0(x) \exp (-x)$ represents the leading
angular dependence.  The last integral cannot be recast in a simple
form and must be evaluated numerically (it also depends on the
cosmological model).  We stress that, since $\vec\delta$ in
Eq.~\eqref{eq:81} has been taken to be expressed in radians, we
similarly need to express $\sigma_W$ in radians in Eq.~\eqref{eq:90}.

\begin{figure}[!t]
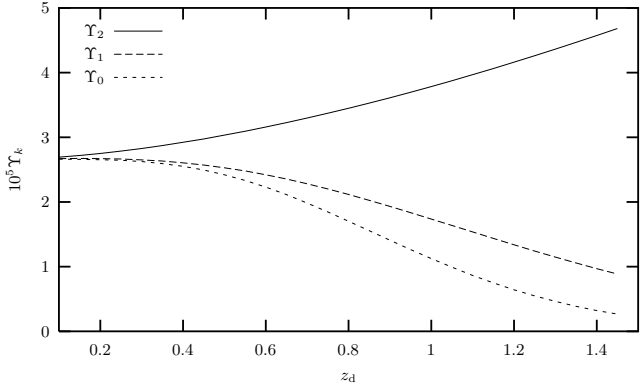

  \begin{center}
    \resizebox{\hsize}{!}{\input fig4.tex}
    \caption{The dimensionless quantities $\Upsilon_k$ as a
      function of the lens redshift $z_\mathrm{d}$.}
    \label{fig:8}
  \end{center}
\end{figure}

Calculations for the other integrals involving the two-point
correlation function, namely $\Xi_i$ and $\Xi$, are very similar.
Defining the dimensionless quantities
\begin{equation}
  \label{eq:91}
  \Upsilon_k = \frac{\pi^{3/2} d_0^2}{2 \langle Z \rangle^k}
  \int_{z_d}^\infty \frac{(1 + z)^\alpha \bigl[p_z(z) \bigr]^2 \bigl[
  Z(z) \bigr]^k}{D_A(z) D_P'(z)} \, \diff z \; ,
\end{equation}
the various two-point correlation terms can be written as
\begin{gather}
  \label{eq:92}
  \Xi_{ij} = \frac{\langle Z \rangle^2}{\sigma_W}
  \gamma_i(\vecbar\theta) \gamma_j(\vecbar\theta) \, \eta \left(
    \frac{\lvert \vec\varphi \rvert^2}{8 \sigma_W^2} \right)
  \Upsilon_2 \; , \\
  \label{eq:93}
  \Xi_i = \frac{\langle Z \rangle}{\sigma_W}
  \gamma_i(\vecbar\theta) \, \eta \left( \frac{\lvert \vec\varphi
  \rvert^2}{8 \sigma_W^2} \right) \Upsilon_1 \; , \\
  \label{eq:94}
  \Xi = \frac{1}{\sigma_W} \, \eta \left( \frac{\lvert \vec\varphi
      \rvert^2}{8 \sigma_W^2} \right) \Upsilon_0 \; .
\end{gather}
In summary, the contribution to the $\Cov(\hat\gamma)$ from a
non-vanishing two-point correlation function is
\begin{align}
  \label{eq:95}
  C_2 = \frac{1}{\sigma_W} \Bigl[ & \Upsilon_2 \gamma_i(\vecbar\theta)
  \gamma_j(\vecbar\theta) + \Upsilon_0 \gamma_i(\vec\theta)
  \gamma_j(\vec\theta') - \Upsilon_1
  \gamma_i(\vec\theta) \gamma_j(\vecbar\theta) \notag\\
  & {} - \Upsilon_1 \gamma_i(\vecbar\theta) \gamma_j(\vec\theta')
  \Bigr] \eta \left( \frac{\lvert \vec\theta - \vec\theta' \rvert^2}{8
      \sigma_W^2} \right) \; .
\end{align}
Note that $\eta(x) \propto 1/\sqrt{x}$ for large $x$, and thus this
contribution to the covariance is the leading term at large
separations $\lvert \vec\phi \rvert$.

In the case $\vec\theta = \vec\theta'$ this expression can be written
in the more compact form
\begin{equation}
  \label{eq:96}
  C_2 = \frac{\Upsilon_2 + \Upsilon_0 - 2 \Upsilon_1}{\sigma_W}
  \gamma_i(\vec\theta) \gamma_j(\vec\theta) \; .
\end{equation}
For a lens at redshift $z_\mathrm{d} = 0.3$ we have $\Upsilon_2 \simeq
2.8267 \cdot 10^{-5}$, $\Upsilon_1 \simeq 2.6513 \cdot 10^{-5}$, and
$\Upsilon_0 \simeq 2.6246 \cdot 10^{-5}$, so that $\Upsilon_2 +
\Upsilon_0 - 2 \Upsilon_1 \simeq 1.488 \cdot 10^{-6}$.  A plot of the
quantities $\Upsilon_k$ as a function of the lens redshift is provided
in Fig.~\ref{fig:8}.

For Eq.~\eqref{eq:4} we obtain similar results, but now only the term
$\Xi_{ij}$ contributes to the covariance.  More specifically, we find
\begin{equation}
  \label{eq:97}
  C_2 = \frac{\Upsilon_2}{\sigma_W} \gamma_i(\vecbar\theta)
  \gamma_j(\vecbar\theta) \eta \left( \frac{\lvert \vec\theta -
      \vec\theta' \rvert^2}{8 \sigma_W^2} \right) \; .
\end{equation}
As a result, we get a much larger error, since the effect of
cancellation of terms clearly shown by Eq.~\eqref{eq:96} is no longer
present.  In particular, for the typical case considered here, we
expect a correlation about $19$ times larger, i.e. an error about
$4$ times larger.  

\section{Calculations for estimators with source redshifts}
\label{sec:calc-estim-with}

In this appendix we briefly sketch the calculations leading to the
results described in Sect.~\ref{sec:case-source-with}.

\subsection{Unweighted estimator}
\label{sec:unweighted-estimator-1}

Let us call $N$ and $D$ the numerator and the denominator of the
r.h.s.\ of Eq.~\eqref{eq:56}, respectively.  Assuming unbiased
measured cosmological weights, we easily find
\begin{gather}
  \label{eq:98}
  \langle N \rangle = \rho \langle Z \rangle \int_\Omega
  W(\vec\theta - \vec\theta') \gamma(\vec\theta') \, \diff^2 \theta'
  \; , \\
  \label{eq:99}
  \langle D \rangle = \rho \langle Z \rangle \; ,
\end{gather}
so that Eq.~\eqref{eq:56} is recovered.

Regarding the covariance, we find after some calculations
\begin{align}
  \label{eq:100}
  & \Cov (N_i D'; \vec\theta, \vec\theta') = \notag\\
  &{} \bigl\langle Z^2 \bigr\rangle \rho \int_\Omega W(\vec\theta - \vec\phi)
  W(\vec\theta' - \vec\phi) \gamma_i(\vec\phi) \, \diff^2
  \phi \notag\\
  &\qquad {} + \rho^2 \Xi\bigl[ W \gamma_i Z W' Z'
  \bigr] \; , \\
  \label{eq:101}
  & \Cov(D D'; \vec\theta, \vec\theta') = \bigl\langle \hat Z^2
  \bigr\rangle \rho \int_\Omega W(\vec\theta - \vec\phi) W(\vec\theta' -
  \vec\phi) \, \diff^2 \phi \notag\\
  & \qquad {} + \rho^2 \Xi\bigl[ W Z W' Z' \bigr] \; .
\end{align}
Note that, since the numerator is formally identical to the numerator
of Eq.~\eqref{eq:5}, Eq.~\eqref{eq:73} still holds.  Differences with
respect to the analogous terms calculated above in Eqs.~\eqref{eq:74}
and \eqref{eq:75} are the presence of $\bigl\langle Z^2 \bigr\rangle$
and $\bigl\langle \hat Z^2 \bigr\rangle$ factors instead of $\langle Z
\rangle^2$ and the different form of the correlation terms, which now
include always the redshift $Z$.  Substituting these expressions in
Eq.~\eqref{eq:44} and using the slow varying shear approximation we
obtain Eq.~\eqref{eq:57}.

\subsection{Weighted estimator}
\label{sec:weighted-estimator-1}

The slowly varying shear approximation applied to the numerator $N$
and denominator $D$ of Eq.~\eqref{eq:58} gives
\begin{gather}
  \label{eq:102}
  \langle N \rangle = \rho \langle Z \rangle \int_\Omega
  W(\vec\theta - \vec\theta') \gamma(\vec\theta') \, \diff^2 \theta'
  \; , \\
  \label{eq:103}
  \langle D \rangle = \rho \langle Z \rangle \; .
\end{gather}
These equations lead immediately to Eq.~\eqref{eq:59}. 

Regarding the covariances, we find
\begin{align}
  \label{eq:104}
  & \Cov(N_i N'_j; \vec\theta, \vec\theta') = \sigma^2_\epsilon
  \bigl\langle \hat Z^2 \bigr\rangle \delta_{ij} \int_\Omega
  W(\vec\theta - \vec\theta'') W(\vec\theta' - \vec\theta'') \,
  \diff^2 \theta'' \notag\\
  & \qquad {} + \bigl\langle Z^2 \hat Z^2 \bigr\rangle \rho \int_\Omega
  \gamma_i(\vec\theta'') \gamma_j(\vec\theta'') W(\vec\theta -
  \vec\theta'') W(\vec\theta' - \vec\theta'') \, \diff^2 \theta'' \notag\\
  & \qquad {} + \rho^2 \Xi\bigl[ W'' \gamma_i'' Z'' \hat Z'' W'''
  \gamma_j''' Z''' \hat Z''' \bigr] \; , \\
  \label{eq:105}
  & \Cov (N_i D'; \vec\theta, \vec\theta') = \notag\\
  & \qquad {} + \bigl\langle Z \hat Z^3 \bigr\rangle \rho \int_\Omega
  W(\vec\theta - \vec\theta'') W(\vec\theta' - \vec\theta'')
  \gamma_i(\vec\theta'') \, \diff^2 \theta'' \notag\\
  & \qquad {} + \rho^2 \Xi\bigl[ W'' \gamma_i'' Z'' \hat Z'' (\hat Z''')^2
  W''' \bigr] \; , \\
  \label{eq:106}
  & \Cov (D D'; \vec\theta, \vec\theta') = \bigl\langle \hat Z^4
  \bigr\rangle \rho \int_\Omega W(\vec\theta - \vec\theta'')
  W(\vec\theta' - \vec\theta'') \, \diff^2 \theta'' \notag\\
  & \qquad {} + \rho^2 \Xi\bigl[ W'' (\hat Z'')^2 W''' (\hat Z''')^2
  \bigr] \; .
\end{align}
These expressions can be used to derive Eq.~\eqref{eq:60}.

\newcommand{\apj}[0]{ApJ}
\newcommand{\apjl}[0]{ApJ}
\newcommand{\aj}[0]{AJ}
\newcommand{\aap}[0]{A\&A}
\newcommand{\aaps}[0]{A\&AS}
\newcommand{\apjs}[0]{ApJSS}
\newcommand{\mnras}[0]{MNRAS}
\newcommand{\pasp}[0]{PASP}
\newcommand{\nat}[0]{Nature}
\newcommand{\araa}[0]{ARA\&A}
\bibliographystyle{apj}
\bibliography{lens-refs} 

\begin{thebibliography}{42}
\expandafter\ifx\csname natexlab\endcsname\relax\def\natexlab#1{#1}\fi

\bibitem[{Bartelmann {et~al.}(1996)Bartelmann, Narayan, Seitz, \&
  Schneider}]{1996ApJ...464L.115B}
Bartelmann, M., Narayan, R., Seitz, S., \& Schneider, P. 1996, \apjl, 464, L115

\bibitem[{{Bartelmann} \& {Schneider}(2001)}]{RevBS}
{Bartelmann}, M. \& {Schneider}, P. 2001, {Physics} {Reports}, 340, 291

\bibitem[{{Benitez}(2000)}]{2000ApJ...536..571B}
{Benitez}, N. 2000, \apj, 536, 571

\bibitem[{{Bertin} \& {Lombardi}(2001)}]{2001ApJ...546...47B}
{Bertin}, G. \& {Lombardi}, M. 2001, \apj, 546, 47

\bibitem[{Brainerd {et~al.}(1996)Brainerd, Blandford, \&
  Smail}]{1996ApJ...466..623B}
Brainerd, T.~G., Blandford, R.~D., \& Smail, I. 1996, \apj, 466, 623

\bibitem[{{Clowe} {et~al.}(1998){Clowe}, {Luppino}, {Kaiser}, {Henry}, \&
  {Gioia}}]{1998ApJ...497L..61C}
{Clowe}, D., {Luppino}, G., {Kaiser}, N., {Henry}, J., \& {Gioia}, I. 1998,
  \apjl, 497, L61

\bibitem[{Eadie {et~al.}(1971)Eadie, Drijard, James, Roos, \& Sadoulet}]{Eadie}
Eadie, W., Drijard, D., James, F., Roos, M., \& Sadoulet, B. 1971, Statistical
  Methods in Experimental Physics (Amsterdam {New-York} Oxford: {North-Holland}
  Publishing Company)

\bibitem[{{Fahlman} {et~al.}(1994){Fahlman}, {Kaiser}, {Squires}, \&
  {Woods}}]{1994ApJ...437...56F}
{Fahlman}, G., {Kaiser}, N., {Squires}, G., \& {Woods}, D. 1994, \apj, 437, 56

\bibitem[{{Fischer}(1999)}]{1999AJ....117.2024F}
{Fischer}, P. 1999, \aj, 117, 2024

\bibitem[{{Fischer} {et~al.}(1997){Fischer}, {Bernstein}, {Rhee}, \&
  {Tyson}}]{1997AJ....113..521F}
{Fischer}, P., {Bernstein}, G., {Rhee}, G., \& {Tyson}, J.~A. 1997, \aj, 113,
  521

\bibitem[{{Fischer} \& {Tyson}(1997)}]{1997AJ....114...14F}
{Fischer}, P. \& {Tyson}, J.~A. 1997, \aj, 114, 14

\bibitem[{Hoekstra {et~al.}(1998)Hoekstra, Franx, Kuijken, \&
  Squires}]{1998ApJ...504..636H}
Hoekstra, H., Franx, M., Kuijken, K., \& Squires, G. 1998, \apj, 504, 636

\bibitem[{Kaiser(1995)}]{1995ApJ...439L...1K}
Kaiser, N. 1995, \apjl, 439, L1

\bibitem[{Kaiser \& Squires(1993)}]{KS}
Kaiser, N. \& Squires, G. 1993, \apj, 404, 441

\bibitem[{{Kaiser} {et~al.}(1995){Kaiser}, {Squires}, \&
  {Broadhurst}}]{1995ApJ...449..460K}
{Kaiser}, N., {Squires}, G., \& {Broadhurst}, T. 1995, \apj, 449, 460

\bibitem[{{Kochanek}(1990)}]{1990MNRAS.247..135K}
{Kochanek}, C.~S. 1990, \mnras, 247, 135

\bibitem[{{Le F\`evre} {et~al.}(1996){Le F\`evre}, {Hudon}, {Lilly},
  {Crampton}, {Hammer}, \& {Tresse}}]{1996ApJ...461..534L}
{Le F\`evre}, O., {Hudon}, D., {Lilly}, S.~J., {Crampton}, D., {Hammer}, F., \&
  {Tresse}, L. 1996, \apj, 461, 534

\bibitem[{{Lombardi} \& {Bertin}(1998{\natexlab{a}})}]{1998A&A...335....1L}
{Lombardi}, M. \& {Bertin}, G. 1998{\natexlab{a}}, \aap, 335, 1

\bibitem[{{Lombardi} \& {Bertin}(1998{\natexlab{b}})}]{1998A&A...330..791L}
---. 1998{\natexlab{b}}, \aap, 330, 791

\bibitem[{{Lombardi} \& {Bertin}(1999{\natexlab{a}})}]{1999A&A...348...38L}
---. 1999{\natexlab{a}}, \aap, 348, 38

\bibitem[{{Lombardi} \& {Bertin}(1999{\natexlab{b}})}]{1999A&A...342..337L}
---. 1999{\natexlab{b}}, \aap, 342, 337

\bibitem[{{Lombardi} {et~al.}(2000){Lombardi}, {Rosati}, {Nonino}, {Girardi},
  {Borgani}, \& {Squires}}]{2000A&A...363..401L}
{Lombardi}, M., {Rosati}, P., {Nonino}, M., {Girardi}, M., {Borgani}, S., \&
  {Squires}, G. 2000, \aap, 363, 401

\bibitem[{{Lombardi} \& {Schneider}(2001)}]{2001A&A...373..359L}
{Lombardi}, M. \& {Schneider}, P. 2001, \aap, 373, 359

\bibitem[{{Luppino} \& {Kaiser}(1997)}]{1997ApJ...475...20L}
{Luppino}, G.~A. \& {Kaiser}, N. 1997, \apj, 475, 20

\bibitem[{{Mellier}(1999)}]{1999ARA&A..37..127M}
{Mellier}, Y. 1999, \araa, 37, 127

\bibitem[{Peebles(1993)}]{Peebles}
Peebles, P. J.~E. 1993, Princiles of Physical Cosmology (Princeton: Princeton
  University Press)

\bibitem[{{Schneider}(1995)}]{1995A&A...302..639S}
{Schneider}, P. 1995, \aap, 302, 639

\bibitem[{Schneider \& Seitz(1995)}]{SS1}
Schneider, P. \& Seitz, C. 1995, \aap, 294, 411

\bibitem[{Seitz {et~al.}(1996)Seitz, Kneib, Schneider, \&
  Seitz}]{1996A&A...314..707S}
Seitz, C., Kneib, J.-P., Schneider, P., \& Seitz, S. 1996, \aap, 314, 707

\bibitem[{Seitz \& Schneider(1995)}]{SS2}
Seitz, C. \& Schneider, P. 1995, \aap, 297, 287

\bibitem[{Seitz \& Schneider(1997)}]{SS3}
---. 1997, \aap, 318, 687

\bibitem[{Seitz \& Schneider(1996)}]{1996A&A...305..383S}
Seitz, S. \& Schneider, P. 1996, \aap, 305, 383

\bibitem[{{Seitz} \& {Schneider}(2001)}]{2001A&A...374..740S}
{Seitz}, S. \& {Schneider}, P. 2001, \aap, 374, 740

\bibitem[{Seitz {et~al.}(1998)Seitz, Schneider, \&
  Bartelmann}]{1998A&A...337..325S}
Seitz, S., Schneider, P., \& Bartelmann, M. 1998, \aap, 337, 325

\bibitem[{{Smail} {et~al.}(1995){Smail}, {Couch}, {Ellis}, \&
  {Sharples}}]{1995ApJ...440..501S}
{Smail}, I., {Couch}, W., {Ellis}, R., \& {Sharples}, R. 1995, \apj, 440, 501

\bibitem[{Squires \& Kaiser(1996)}]{1996ApJ...473...65S}
Squires, G. \& Kaiser, N. 1996, \apj, 473, 65

\bibitem[{{Squires} {et~al.}(1996){Squires}, {Kaiser}, {Babul}, {Fahlman},
  {Woods}, {Neumann}, \& {Boehringer}}]{1996ApJ...461..572S}
{Squires}, G., {Kaiser}, N., {Babul}, A., {Fahlman}, G., {Woods}, D.,
  {Neumann}, D., \& {Boehringer}, H. 1996, \apj, 461, 572

\bibitem[{Squires {et~al.}(1996)Squires, Kaiser, Fahlman, Babul, \&
  Woods}]{1996ApJ...469...73S}
Squires, G., Kaiser, N., Fahlman, G., Babul, A., \& Woods, D. 1996, \apj, 469,
  73

\bibitem[{{Tyson} {et~al.}(1990){Tyson}, {Wenk}, \&
  {Valdes}}]{1990ApJ...349L...1T}
{Tyson}, J., {Wenk}, R., \& {Valdes}, F. 1990, \apjl, 349, L1

\bibitem[{{Tyson} \& {Fischer}(1995)}]{1995ApJ...446L..55T}
{Tyson}, J.~A. \& {Fischer}, P. 1995, \apjl, 446, L55

\bibitem[{van {Waerbeke}(2000)}]{2000MNRAS.313..524W}
van {Waerbeke}, L. 2000, \mnras, 313, 524

\bibitem[{{Webster}(1985)}]{1985MNRAS.213..871W}
{Webster}, R. 1985, \mnras, 213, 871

\end{thebibliography}

\end{document}